# Stress Softening Damage in Strongly Nonlinear Viscoelastic Soft Materials A Physics-Informed Data-Driven Constitutive Model with Time–Temperature Coupling

Alireza Ostadrahimi[1], Amir Teimouri[1], Kshitiz Upadhyay[2], Guoqiang Li[1]

[1]Department of Mechanical and Industrial Engineering, Louisiana State University, Baton Rouge, LA, USA

[2]Department of Aerospace Engineering and Mechanics, University of Minnesota, Minneapolis, MN, USA

**Abstract**

This study presents a novel physics-informed, data-driven modeling framework for capturing the strongly nonlinear thermo-viscoelastic behavior of soft materials exhibiting stress-softening, with emphasis on the Mullins effect. Unlike previous approaches limited to quasi-static or isothermal conditions, our model unifies rate dependence, temperature sensitivity, large-strain cyclic loading, and evolving damage mechanisms. Thermodynamic admissibility is ensured via a custom loss function that embeds the Clausius–Duhem inequality and explicitly constrains the damage variable for physically realistic softening. A Temporal Convolutional Network is trained on high-fidelity experimental data across multiple temperatures, strain rates, and stretch levels, enabling the model to capture rich thermomechanical coupling and history dependence. The framework generalizes to unseen thermo-mechanical conditions, higher strain rates, and larger deformations, and remains robust to input noise. Validation against finite element simulations using Abaqus/Explicit demonstrates excellent agreement under cyclic loading and damage evolution, confirming the surrogate model's effectiveness for advanced simulation workflows.

## 1   Introduction

Soft materials such as elastomers, biological tissues, and polymer foams demonstrate complex mechanical responses involving large deformations, nonlinear stress-strain relationships, and time-dependent viscoelastic behavior influenced by loading history ([1], [2], [3], [4], [5], [6], [7], [8], [9], [10], [11], [12], [13]). These materials typically undergo a softening response following the initial loading cycle, called the Mullins effect. In composites and rubber-like substances, this phenomenon is often attributed to the breakdown of polymer-filler interfaces, disintegration of filler networks, molecular slippage, and chain disentanglement ([14], [15], [16], [17], [18]). The stress softening effect typically emerges in the rubbery regime but tends to disappear at lower temperatures or when the material is glassy ([19], [20], [21], [22], [23], [24]). A number of researchers have employed strain energy-based continuum frameworks to investigate Mullins effect along with the nonlinear responses characteristic of viscoelastic and hyperelastic materials ([25], [26], [27]). To investigate softening behavior up to 20% strain, classical constitutive frameworks were employed, such as linear viscoelasticity, finite linear viscoelasticity, and quasi-linear viscoelasticity, incorporating the Boltzmann superposition integral and using the Prony series to characterize time-dependent phenomena. These models describe the viscoelastic behavior of soft materials within a linear regime; however, they are inadequate for capturing the nonlinear stress–strain response, strain-dependent relaxation, and softening effects that emerge at larger deformations ([28], [29], [30], [31], [32], [33]).

---
[1] lguoqi1@lsu.edu,
First author email: Ostadrahimi.86@gmail.com

To model materials with finite deformations much larger than 20% strain, various finite nonlinear viscoelastic theories have been formulated. These approaches generally link the material's time-dependent behavior to irreversible thermodynamic processes driven by microstructural evolution. As a result, they incorporate internal or external thermodynamic state variables to capture the rate- and time-dependent mechanical responses ([34], [35], [36], [3], [37], [38], [39], [40], [41], [42], [43], [44], [45], [46]). In particular, internal state variable-based models characterize the non-equilibrium component of the Helmholtz free energy using a set of evolving internal variables ([47], [48], [4], [49], [50], [51], [52]). Flow rules are often used to model the nonlinear behavior observed in viscoelastic and inelastic materials under large deformations. Although nonlinear flow rules are typically phenomenological, they can gain physical relevance when carefully constructed, particularly in capturing stress-driven reductions in relaxation time. However, these models often lack a direct thermodynamic foundation, such as derivation from the Clausius–Duhem inequality ([53], [54], [55], [56]). Addressing these shortcomings, Ostadrahimi et al. [57] building on Holzapfel's (2010) approach, introduced a nonlinear viscous formulation within a multi-branch Maxwell framework to enhance the modeling of shape memory polymers beyond the limitations of linear viscoelastic theories.

Conventional constitutive models often face challenges in capturing multi-physics phenomena, necessitating intricate modifications to account for thermo-mechanical coupling and damage evolution (, [58], [59]). In contrast, Machine Learning (ML) frameworks can inherently learn such dependencies from multi-scale datasets, enhancing predictive capability and computational efficiency ([60], [61]). In the absence of physical constraints, purely data-driven models may yield unrealistic stress responses that breach essential principles like objectivity and thermodynamic consistency. ([62], [63]). Physics-informed approaches such as thermodynamically consistent networks enhance both predictive accuracy and adherence to fundamental physical laws ([64], [65], [66], [67], [68]).

Constitutive Artificial Neural Networks (CANNs), Physics-Informed Neural Networks (PINNs), Pretrained Audio Neural Networks (PANNs), and Bayesian Neural Networks (BNNs) have been employed to impose physical laws for viscoelastic deformation and to quantify uncertainty ([67], [69], [70], [71], [72], [73], [74], [75], [76], [77], [78], [79], [80], [81]). Further developments include Finite Element-based Artificial Neural Networks (FEANNs), Gaussian Process Regression (GPR), and Recurrent Neural Networks (RNNs) with Long Short-Term Memory (LSTM) units, which have proven effective in capturing history-, and rate-dependent stress-strain behavior ([65], [82], [83], [84], [85], [86], [87], [88], [89], [90], [91]). Recent studies have shown that convolutional architectures, particularly Temporal Convolutional Networks (TCNs), consistently outperform recurrent models like LSTM and GRU in both accuracy and memory retention ([92], [93]). TCNs offer a simpler structure with longer effective memory, enabled by dilated convolutions, and have demonstrated superior performance across diverse benchmark tasks ([94], [95], [96], [97], [98]). Several ML-based models have also been successfully applied in plasticity, fatigue, and fracture modeling to capture damage with high predictive accuracy and interpretability. ([99], [100], [101], [102], [103], [104], [105], [106], [107], [108], [109], [110], [111], [112], [113], [114], [115], [116], [117], [118], [119], [120], [121], [122], [123], [124], [125], [126], [127], [128], [129], [130], [131], [132], [133], [134], [135],[136], [137], [138], [139], [140], [141]).

More specifically, regarding Mullins effect in soft materials, Abdusalamov et al. [126] introduced an approach using Deep Symbolic Regression (DSR) to rediscover the softening effect in rubber-like materials by identifying both the hyperelastic strain energy function and a damage evolution function. While their approach remains limited to quasi-static conditions, without time-dependence, and dissipated energy constraints. Recently, Zlatić and Čanađija [108] introduced a physics-augmented neural network framework to recover the Mullins effect in incompressible hyperelastic materials. However, their approach captures stress softening due to isotropic damage, it remains limited to rate-independence and isothermal scenarios.

Therefore, to fill the gap left by existing approaches that neglect rate dependence, temperature effects, dissipative constraints, and the ability to model giant elongations up to 200%, the present study focuses on developing a broadly-applicable modeling framework for soft materials, addressing: (1) extremely nonlinear thermoviscoelastic behavior, (2) time-dependent effects, (3) temperature dependence, (4) Mullins effect, (5) large cyclic elongations up to 200%, (6) strict enforcement of the Clausius-Duhem inequality within the loss function, and (7) physically constraining the evolution of the damage. In contrast to symbolic regression or static neural models, our time-sequential approach employs TCNs, trained against ground truth data from multi-cycle thermomechanical tests conducted at multiple strain rates and temperatures.

Beyond its in-domain performance, the model demonstrates strong generalization capability under previously unseen conditions, accurately predicting responses at temperatures not included in training (e.g., 10°C), strain rates 2.5× higher than those used for training (from 200%/min to 500%/min), and a 50% increase in elongation (from 150% to 200%), as well as longer loading histories (tested on three cycles after being trained on two). This model not only captures stress evolution with high fidelity but is directly implementable into finite element solvers via a VUMAT subroutine, enabling predictive simulations in real-world applications. Thermodynamic consistency is maintained by ensuring that the predicted internal dissipation is strictly non-negative and increases with loading cycles and strain amplitudes. Moreover, robustness under uncertainty was demonstrated by injecting 20% Gaussian noise into the stress input, without loss of predictive accuracy. Finally, the surrogate model's predictions were validated against finite element simulations of cyclic Mullins damage in an open-hole specimen, demonstrating close agreement and high practical relevance.

## 2 Constitutive framework for generalized internal state variable-based thermo-visco-hyperelasticity with Mullins effect

### 2.1 Basic Kinematics

A soft material undergoes a transformation from its original, undeformed state at time *t=0* to a deformed state at a later time *t*. This macroscopic deformation is described by a mapping from the reference position vector **X** to the current spatial position vector **x**, defined through a time-dependent vector function. The associated deformation gradient tensor **F**=∂**x**/∂**X** captures this transformation, with its determinant *J* representing the local volume change.

In a Lagrangian description, deformation is often analyzed using the symmetric right Cauchy-Green deformation tensor $\mathbf{C} = \mathbf{F}(\mathbf{X},t)^\mathrm{T}\mathbf{F}(\mathbf{X},t)$. To distinguish between volume-changing and shape-changing effects in soft materials, the deformation gradient **F** can be multiplicatively decomposed ($\mathbf{F} = (J^{1/3}\mathbf{I})\overline{\mathbf{F}}$) into a dilational part $J^{1/3}\mathbf{I}$ ($J = \det(\mathbf{F})$) and a distortional part $\overline{\mathbf{F}}$. This leads to the definition of a modified right Cauchy-Green tensor, where $\overline{\mathbf{C}} = J^{-2/3}\mathbf{C}$, removing the volumetric component from **C**.

To describe the material response more systematically, the Helmholtz free energy density $\Psi$, defined per unit reference volume, is split into two contributions: one capturing volumetric effects $\Psi_{\mathrm{VOL}}^{\infty}(J,\Theta)$ (superscript ∞ signifies mechanical equilibrium, and θ denotes temperature) and the other accounting for isochoric (distortional) effects. Within the generalized visco-hyperelastic framework based on internal state variables, the isochoric portion of the free energy is further divided into an equilibrium term representing the long-term thermodynamic response, and a set of configurational energy terms that characterize the time-dependent (non-equilibrium) behavior of the material. These non-equilibrium contributions, denoted as scalar functions $\Upsilon_\alpha, \alpha = 1,...,m$, depend on internal variables $\mathbf{\Gamma}_\alpha$. Altogether, the total Helmholtz free energy comprises both volumetric and isochoric components, encompassing equilibrium and non-equilibrium contributions.

$$\Psi(\mathbf{C},\mathbf{\Gamma}_1,...,\mathbf{\Gamma}_m,\Theta) = \Psi^\infty_{\text{VOL}}(J,\Theta) + \Psi^\infty_{\text{ISO}}(\overline{\mathbf{C}},\Theta) + \sum_{\alpha=1}^{m} \Upsilon_\alpha(\overline{\mathbf{C}},\mathbf{\Gamma}_\alpha,\Theta) \tag{1}$$

The temperature dependence of soft materials is a critical and inherently complex factor that governs their mechanical performance, processing behavior, and chemical stability. Due to the strong coupling between time, temperature, and molecular relaxation dynamics arising from chain rearrangements, the explicit temperature dependence in **Eq. 1** is omitted in the rest of this paper. Instead, the Time-Temperature Superposition (TTS) technique is adopted as a more robust and physically grounded approach to capture the temperature-induced shifts in relaxation times. This enables a precise and unified description of the soft materials' viscous behavior across a wide range of thermal conditions. To capture the relaxational behavior of multiple polymer chains, the model incorporates $m$ irreversible terms, each linked to a characteristic relaxation time $\tau_\alpha \in (0,\infty)$.

The strain energy function in **Eq. 1** is thereby extended through a formulation that accounts for stress-softening and damage evolution. Notably, internal damage and microstructural reconfiguration tend to degrade the material's shear resistance far more significantly than its volumetric response [142]. As damage accumulates, the material undergoes stiffness reduction and structural changes that directly impact the configurational free energy landscape. To account for this, the Helmholtz free energy is redefined by considering $^{\text{intact}}\Psi^\infty_{\text{ISO}}(\overline{\mathbf{C}})$ and $\sum_{\alpha=1}^{m} {}^{\text{intact}}\Upsilon_\alpha(\overline{\mathbf{C}},\mathbf{\Gamma}_\alpha)$ as the *undamaged* (indicated by subscript 'intact') equilibrium isochoric free energy component and the non-equilibrium configurational energy component, respectively.

Introducing an isotropic damage variable $\vartheta \in [0,1]$ to capture Mullins' effect, along with the corresponding integrity factor $(1-\vartheta)$, allows for a direct modification of the strain energy function. This yields a final expression that robustly captures the effects of stress-softening damage on the mechanical and energetic response of the soft materials network.

$$\Psi(\mathbf{C},\mathbf{\Gamma}_1,...,\mathbf{\Gamma}_m,\vartheta) = \Psi^\infty_{\text{VOL}}(J) + (1-\vartheta)\left({}^{\text{intact}}\Psi^\infty_{\text{ISO}}(\overline{\mathbf{C}}) + \sum_{\alpha=1}^{m} {}^{\text{intact}}\Upsilon_\alpha(\overline{\mathbf{C}},\mathbf{\Gamma}_\alpha)\right) \tag{2}$$

The isotropic damage variable can evolve with deformation and time, with $\vartheta = 1$ signifying no damage and $\vartheta = 0$ signifying complete mechanical damage. $\Psi^\infty_{\text{VOL}}(1) = 0$, ${}^{\text{intact}}\Psi^\infty_{\text{ISO}}(\mathbf{I}) = 0$, and $\sum_{\alpha=1}^{m} {}^{\text{intact}}\Upsilon_\alpha(\mathbf{I},\mathbf{I}) = 0$ are the normalization conditions. The Clausius-Duhem inequality, a cornerstone of thermodynamics, embodies the second law and serves as the basis for quantifying internal dissipation $D_{\text{int}}$ during deformation processes. Under isothermal conditions, and when working with the second Piola-Kirchhoff stress tensor $\mathbf{S}$, the internal dissipation can be expressed as

$$D_{int} = \mathbf{S} : \dot{\mathbf{C}}/2 - \dot{\Psi} \geq 0 \tag{3}$$

where $\dot{\mathbf{C}}$ is the material time derivative of $\mathbf{C}$. In this formulation, $\mathbf{S}$ and $\dot{\mathbf{C}}$ are thermodynamically conjugated as determined via the Coleman–Noll procedure, which imposes thermodynamic consistency on the constitutive equations. For compressible hyperelastic materials, $D_{int} \geq 0$ inequality must be satisfied, and in purely elastic (non-dissipative) cases, this reduces to an equality ($D_{int} = 0$). Using the free energy expression in **Eq. 2** and applying chain rule, the rate of change of free energy becomes:

$$\dot{\Psi} = \frac{\partial \Psi_{VOL}^{\infty}(J)}{\partial J} \dot{J} - \dot{\vartheta}\left( {}^{intact}\Psi_{ISO}^{\infty}(\overline{\mathbf{C}}) + \sum_{\alpha=1}^{m} {}^{intact}\Upsilon_{\alpha}(\overline{\mathbf{C}}, \Gamma_{\alpha}) \right)$$
$$+ (1-\vartheta)\left( \frac{\partial {}^{intact}\Psi_{ISO}^{\infty}(\overline{\mathbf{C}})}{\partial \overline{\mathbf{C}}} + \sum_{\alpha=1}^{m} \frac{\partial {}^{intact}\Upsilon_{\alpha}(\overline{\mathbf{C}}, \Gamma_{\alpha})}{\partial \overline{\mathbf{C}}} \right) : \dot{\overline{\mathbf{C}}} + \sum_{\alpha=1}^{m} \frac{\partial {}^{intact}\Upsilon_{\alpha}(\overline{\mathbf{C}}, \Gamma_{\alpha})}{\partial \Gamma_{\alpha}} : \dot{\Gamma}_{\alpha}(1-\vartheta) \tag{4}$$

The derivatives of free energy components with respect to $\overline{\mathbf{C}}$ can be transformed into corresponding derivatives with respect to $\mathbf{C}$ using chain rule. Further, considering the time derivative $\dot{\overline{\mathbf{C}}}$ ($\dot{\overline{\mathbf{C}}} = \partial \overline{\mathbf{C}}/\partial \mathbf{C} : \dot{\mathbf{C}}$) as well as $\dot{J} = \partial J/\partial \mathbf{C} : \dot{\mathbf{C}} = J\mathbf{C}^{-1} : \dot{\mathbf{C}}/2$, the internal dissipation can be reformulated as:

$$D_{int} = \left[ \mathbf{S} - J\frac{\partial \Psi_{VOL}^{\infty}(J)}{\partial J}\mathbf{C}^{-1} - 2(1-\vartheta)\left( \frac{\partial {}^{intact}\Psi_{ISO}^{\infty}(\overline{\mathbf{C}})}{\partial \mathbf{C}} + \sum_{\alpha=1}^{m} \frac{\partial {}^{intact}\Upsilon_{\alpha}(\overline{\mathbf{C}}, \Gamma_{\alpha})}{\partial \mathbf{C}} \right) \right] : \frac{1}{2}\dot{\mathbf{C}}$$
$$- (1-\vartheta)\sum_{\alpha=1}^{m} \frac{\partial {}^{intact}\Upsilon_{\alpha}(\overline{\mathbf{C}}, \Gamma_{\alpha})}{\partial \Gamma_{\alpha}} : \dot{\Gamma}_{\alpha} + \dot{\vartheta}\left( {}^{intact}\Psi_{ISO}^{\infty}(\overline{\mathbf{C}}) + \sum_{\alpha=1}^{m} {}^{intact}\Upsilon_{\alpha}(\overline{\mathbf{C}}, \Gamma_{\alpha}) \right) \geq 0 \tag{5}$$

Following $\mathbf{S} = 2\nabla_{\mathbf{C}}\Psi$ the total stress can be systematically decomposed into three distinct contributions: a volumetric equilibrium component $\mathbf{S}_{VOL}^{\infty}$, an isochoric hyperelastic equilibrium component $\mathbf{S}_{ISO}^{\infty}$ and a non-equilibrium (viscous or dissipative) component $\sum_{\alpha=1}^{m} \mathbf{Q}_{\alpha}$. Accordingly, the total second Piola-Kirchhoff stress is expressed as:

$$\mathbf{S} = \mathbf{S}_{VOL}^{\infty} + \mathbf{S}_{ISO}^{\infty} + \sum_{\alpha=1}^{m} \mathbf{Q}_{\alpha} = J\frac{\partial \Psi_{VOL}^{\infty}(J)}{\partial J}\mathbf{C}^{-1} + 2(1-\vartheta)\frac{\partial {}^{intact}\Psi_{ISO}^{\infty}(\overline{\mathbf{C}})}{\partial \mathbf{C}} + (1-\vartheta)\sum_{\alpha=1}^{m} 2\frac{\partial {}^{intact}\Upsilon_{\alpha}(\overline{\mathbf{C}}, \Gamma_{\alpha})}{\partial \mathbf{C}} \tag{6}$$

The equilibrium components $\mathbf{S}_{\text{VOL}}^{\infty}$ and $\mathbf{S}_{\text{ISO}}^{\infty}$ are strictly derived from thermodynamic potentials and represent purely recoverable elastic effects. These stresses are reversible, do not depend on the deformation history, and do not contribute to dissipation; their role is confined to storing mechanical energy. In contrast, the only source of entropy production originates from the non-equilibrium configurational stress components, denoted by $\sum_{\alpha=1}^{m} \mathbf{Q}_{\alpha}$. Unlike their equilibrium counterparts, they are inherently irreversible, directly responsible for internal dissipation, and drive the system toward thermodynamic equilibrium over time. This separation provides a rigorous foundation for defining a non-negative internal dissipation inequality. Constructing a thermodynamically consistent constitutive model yield:

$$\begin{cases} \mathbf{S}_{\text{VOL}}^{\infty} = J \dfrac{\partial \Psi_{\text{VOL}}^{\infty}(J)}{\partial J} \mathbf{C}^{-1}, \\[6pt] \mathbf{S}_{\text{ISO}}^{\infty} = 2(1-\vartheta) \dfrac{\partial^{\text{intact}} \Psi_{\text{ISO}}^{\infty}(\overline{\mathbf{C}})}{\partial \mathbf{C}}, \\[6pt] \mathbf{Q}_{\alpha} = 2(1-\vartheta) \dfrac{\partial^{\text{intact}} \Upsilon_{\alpha}(\overline{\mathbf{C}}, \boldsymbol{\Gamma}_{\alpha})}{\partial \mathbf{C}}, \\[6pt] \mathbf{S}_{\text{ISO}} = \mathbf{S}_{\text{ISO}}^{\infty} + \sum_{\alpha=1}^{m} \mathbf{Q}_{\alpha}, \\[6pt] D_{\text{int}} = \dot{\vartheta}\left(^{\text{intact}} \Psi_{\text{ISO}}^{\infty}(\overline{\mathbf{C}}) + \sum_{\alpha=1}^{m} {}^{\text{intact}} \Upsilon_{\alpha}(\overline{\mathbf{C}}, \boldsymbol{\Gamma}_{\alpha})\right) - (1-\vartheta) \sum_{\alpha=1}^{m} \dfrac{\partial^{\text{intact}} \Upsilon_{\alpha}(\overline{\mathbf{C}}, \boldsymbol{\Gamma}_{\alpha})}{\partial \boldsymbol{\Gamma}_{\alpha}} : \dot{\boldsymbol{\Gamma}}_{\alpha} \geq 0 \end{cases} \qquad (7)$$

The non-equilibrium stress components, expressed as $\sum_{\alpha=1}^{m} \mathbf{Q}_{\alpha}$, encapsulate the irreversible kinetic processes governed by a set of tensor-valued internal variables. These variables are intrinsically linked to the non-equilibrium Maxwell branches, which characterize the time-dependent viscoelastic response of the material. Capturing this non-equilibrium behavior necessitates a dual framework: a rheological model and a corresponding evolution equation. The rheological model, such as the generalized Maxwell model in **Figure 1**, accounts for both the immediate elastic response and the delayed, time-dependent viscous effects. Complementing this, the evolution equation rigorously defines the dynamic progression of internal state variables under loading/unloading, thus enabling a precise representation of the material's viscoelastic memory and dissipative behavior.

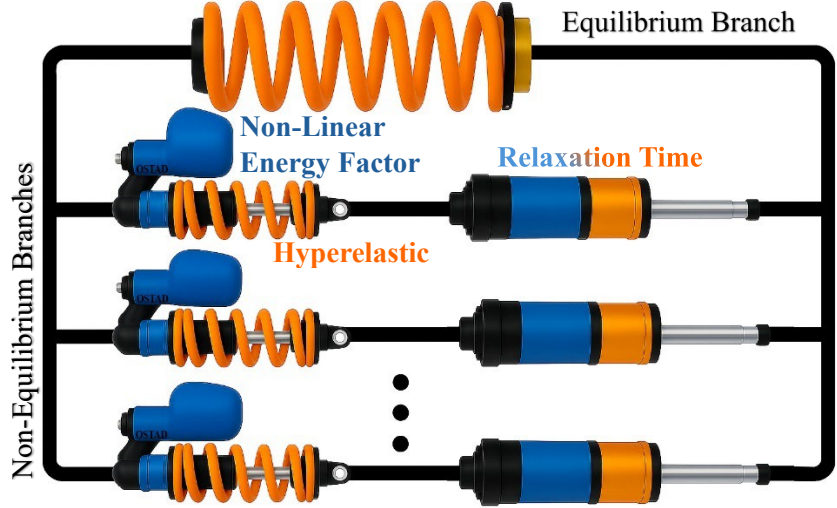

Figure 1: Nonlinear Generalized Maxwell in Viscous Parts

*2.1.1   Equilibrium component*

Considering continuous derivatives of the second Piola–Kirchhoff stress for the equilibrium part, as well as the projection tensor $\mathbb{P}:\mathbf{Z}=\text{Dev}(\mathbf{Z})$ where , and $\frac{\partial \overline{\mathbf{C}}}{\partial \mathbf{C}}=J^{-2/3}\mathbb{P}^T$ where $\mathbb{I}-\frac{1}{3}\mathbf{C}\otimes\mathbf{C}^{-1}$ ([28]) can lead to the total equilibrium isochoric stress component being written as

$$\mathbf{S}^\infty = \frac{\partial \Psi_{\text{VOL}}^\infty(J)}{\partial J}\frac{\partial J}{\partial \mathbf{C}} + 2(1-\vartheta)\sum_{k=1}^{2}\frac{\partial^{\text{intact}}\Psi_{\text{ISO}}^\infty(\overline{\mathbf{C}})}{\partial \overline{I}_k}\frac{\partial \overline{I}_k}{\partial \mathbf{C}} \qquad (8)$$

where $I_k$ (k = 1, 2) are principal invariants of **C**. Using $\left(\frac{\partial \overline{\mathbf{C}}}{\partial \mathbf{C}}\right)^T:\frac{\partial I_k}{\partial \overline{\mathbf{C}}}=J^{-2/3}\text{Dev}\left(\frac{\partial I_k}{\partial \overline{\mathbf{C}}}\right)$ and $\frac{\partial J}{\partial \mathbf{C}}=\frac{1}{2}J\mathbf{C}^{-1}$, as well as the derivative of the first invariant $\frac{\partial I_1}{\partial \mathbf{C}}=\mathbf{I}$ and derivatives of the second invariants $\frac{\partial I_2}{\partial \mathbf{C}}=I_1\mathbf{I}-\mathbf{C}$, **Eq.8** yields:

$$\mathbf{S}^\infty = J\frac{\partial \Psi_{\text{VOL}}^\infty(J)}{\partial J}\mathbf{C}^{-1} \\ + 2J^{-2/3}J^{-2/3}2(1-\vartheta)\left[\frac{\partial^{\text{intact}}\Psi_{\text{ISO}}^\infty(\overline{\mathbf{C}})}{\partial \overline{I}_1}\text{Dev}\left(\frac{\partial \overline{I}_1}{\partial \overline{\mathbf{C}}}\right) + \frac{\partial^{\text{intact}}\Psi_{\text{ISO}}^\infty(\overline{\mathbf{C}})}{\partial \overline{I}_2}\text{Dev}\left(\frac{\partial \overline{I}_2}{\partial \overline{\mathbf{C}}}\right)\right] \qquad (9)$$

*2.1.2   Non-equilibrium component and Mullins Damage Evolution*

The evolution equation provides a rigorous mathematical relationship between stress, strain rate, and deformation history, while strictly complying with the entropy inequality dictated by the second law of thermodynamics. While the Boltzmann superposition principle is traditionally employed to describe viscoelastic behavior, it falls short in capturing the complex nonlinear response of materials. Thus, to overcome these limitations, differential equation-based constitutive frameworks have been employed in this work to capture the evolution of $\mathbf{Q}_\alpha$, offering enhanced fidelity in modeling nonlinear viscoelasticity and dissipative mechanisms ([17], [120]). This form lays the foundation for incorporating irreversible effects into the constitutive behavior of advanced soft materials.

$$\frac{d\mathbf{Q}_\alpha}{dt} = \hat{f}(\overline{\mathbf{C}}, \vartheta, \mathbf{\Gamma}_1, ..., \mathbf{\Gamma}_\alpha) \quad \alpha = 1, ..., m \tag{10}$$

where $\hat{f}$ denotes a constitutive function that dictates the evolution of internal variables. It encapsulates the material-specific kinetics governing irreversible processes such as stress relaxation, creep, and damage evolution while ensuring that the progression of internal variables remains consistent with thermodynamic constraints and path dependency.

The Mullins effect is commonly characterized by stress-softening damage, where material softening accumulates with increasing or repeated deformation. This phenomenon is typically represented through internal variables that evolve based on deformation history, particularly concerning the maximum experienced strain energy. Damage progression is path-dependent and irreversible, often activated only when the current loading exceeds the previous maxima. A general formulation for such damage behavior can be expressed as:

$$\vartheta = \vartheta_\infty \cdot \Im\left(^{intact}\Psi_{ISO,max}, \Psi_{SAT}, \varsigma, {}^{intact}\Psi_{ISO}\right) \tag{11}$$

where $\vartheta_\infty \in [0,1]$ is a scalar damage parameter, $\Psi_{SAT}$ is the saturation level of damage, $^{intact}\Psi_{ISO}$ denotes the current isochoric strain energy in the intact or undamaged configuration, and $^{intact}\Psi_{ISO,max}$ represents its historical maximum over the loading path. The function $\Im(.)$ governs the evolution of damage and may also depend on additional internal variables $\varsigma$, such as the rate of deformation and loading direction. This structure captures the essential features of the Mullins effect, enabling a thermodynamically consistent representation of energy dissipation and stiffness reduction during cyclic or monotonic loading.

### 2.1.3 Temperature Dependency

Within the TTS framework, rescaling the time domain through a shift factor is imperative for capturing the profound influence of temperature on polymer behavior. This is achieved using the Williams-Landel-Ferry (WLF) and/or Arrhenius [143] relations, both of which share a fundamentally equivalent form. The parameters $C_1, C_2$, and $\theta_{ref}$ (or their Arrhenius analogs) are intrinsic to the material, and the reduced time is rigorously defined as:

$$\tau^*(t) = \int_0^t \frac{d\varphi}{A_{T(\varphi)}}$$

$$Log\ A_T = -C_1\left(T - T_{ref}\right)\big/\left(C_2 + \left(T - T_{ref}\right)\right)$$
(12)

where $A_T$ is the shift factor linking temperature shift (concerning a reference temperature $T_{ref}$) and time $t$. Given this, we categorically exclude any explicit temperature dependency in the formulation of the total strain energy function $\Psi$. Instead, we adopt this reduced time framework as a more rigorous and thermodynamically consistent method to capture both the material's viscous response and stress-softening behavior.

## 2.2 An Alternative Representation of the Temperature-Dependent Visco-Hyperelastic Model Including Mullins Effect

It has been shown that the stress components in the generalized internal state variable-based visco-hyperelastic framework can be expressed as a weighted linear combination of elements from a chosen set of irreducible integrity bases. Upadhyay et al. showed that ML techniques can be employed to map strain and strain-rate invariants to the corresponding coefficients, termed response functions, within this basis . This approach inherently satisfies key physical constraints, such as local action, determinism, objectivity, angular momentum balance, and isotropy. By constructing ML models around invariant-response mappings instead of raw stress-strain data, the framework improves data efficiency, enhances generalization to unseen deformation states, and allows for direct comparison with theoretical models by isolating and interpreting learned response functions. Following this approach, we apply the chain rule to expand the generalized stress components in **Eq.7**, resulting in an alternative representation of the generalized internal state variable-based visco-hyperelastic constitutive model, which now explicitly incorporates Mullins damage evolution for a more comprehensive formulation:

$$\mathbf{S} = J\frac{\partial \Psi_{VOL}^{\infty}(J)}{\partial J}\mathbf{C}^{-1}$$

$$+ J^{-2/3}\left[2(1-\vartheta)\left(\frac{\partial^0 \Psi_{ISO}^{\infty}(\overline{\mathbf{C}})}{\partial \overline{I}_1} + \overline{I}_1\frac{\partial^0 \Psi_{ISO}^{\infty}(\overline{\mathbf{C}})}{\partial \overline{I}_2}\right)Dev(\mathbf{I}) - 2(1-\vartheta)\frac{\partial^0 \Psi_{ISO}^{\infty}(\overline{\mathbf{C}})}{\partial \overline{I}_2}Dev(\overline{\mathbf{C}})\right.$$

$$\left.+ 2(1-\vartheta)\overline{I}_3\frac{\partial^0 \Psi_{ISO}^{\infty}(\overline{\mathbf{C}})}{\partial \overline{I}_3}Dev(\overline{\mathbf{C}}^{-1})\right]$$

$$+ J^{-2/3}\left[\sum_{\alpha=1}^{m}2(1-\vartheta)\left(\frac{\partial \Upsilon_\alpha(\overline{\mathbf{C}},\boldsymbol{\Gamma}_\alpha)}{\partial \overline{I}_1} + \overline{I}_1\frac{\partial \Upsilon_\alpha(\overline{\mathbf{C}},\boldsymbol{\Gamma}_\alpha)}{\partial \overline{I}_2}\right)Dev(\mathbf{I}) - \sum_{\alpha=1}^{m}2(1-\vartheta)\frac{\partial \Upsilon_\alpha(\overline{\mathbf{C}},\boldsymbol{\Gamma}_\alpha)}{\partial \overline{I}_2}Dev(\overline{\mathbf{C}})\right.$$

$$\left.+ \sum_{\alpha=1}^{m}2(1-\vartheta)\overline{I}_3\frac{\partial \Upsilon_\alpha(\overline{\mathbf{C}},\boldsymbol{\Gamma}_\alpha)}{\partial \overline{I}_3}Dev(\overline{\mathbf{C}}^{-1})\right]$$
(13)

Based on the above expanded form, the three stress components of the model can be written as weighted linear decompositions of integrity basis tensors as:

$$\mathbf{S}_{\text{VOL}}^{\infty} = \delta(J)\mathbf{C}^{-1} \tag{14}$$

$$\mathbf{S}_{\text{ISO}}^{\infty} = J^{-2/3}(1-\vartheta)\left[\chi_1(\overline{I}_1,\overline{I}_2,\overline{I}_{1,\max})\text{Dev}(\mathbf{I}) + \chi_2(\overline{I}_1,\overline{I}_2,\overline{I}_{1,\max})\text{Dev}(\overline{\mathbf{C}})\right]$$

$$\sum_{\alpha=1}^{m}\mathbf{Q}_\alpha = J^{-2/3}(1-\vartheta)\left[\sum_{\alpha=1}^{m}\xi_{1,\alpha}(t,T,\overline{I}_1,\overline{I}_2,\overline{I}_{1,\max})\text{Dev}(\mathbf{I}) + \sum_{\alpha=1}^{m}\xi_{2,\alpha}(t,T,\overline{I}_1,\overline{I}_2,\overline{I}_{1,\max})\text{Dev}(\overline{\mathbf{C}})\right]$$

Here, the $\mathbb{G}_1 = \mathbf{C}^{-1}$, $\mathbb{G}_2 = \text{Dev}(\mathbf{I})$, and $\mathbb{G}_3 = \text{Dev}(\overline{\mathbf{C}})$ are the three irreducible integrity basis tensors, and $\delta$, $\chi_1$, $\chi_2$, $\xi_{1,\alpha}$ and $\xi_{2,\alpha}$ are scalar response functions that serve as coefficients of the integrity bases:

$$\delta(J) = J\frac{\partial \Psi_{\text{VOL}}^{\infty}(J)}{\partial J}, \tag{15}$$

$$\chi_1(\overline{I}_1,\overline{I}_2,\overline{I}_{1,\max}) = 2\left(\frac{\partial^{\text{intact}}\Psi_{\text{ISO}}^{\infty}(\overline{I}_1,\overline{I}_2)}{\partial \overline{I}_1} + \overline{I}_1\frac{\partial^{\text{intact}}\Psi_{\text{ISO}}^{\infty}(\overline{I}_1,\overline{I}_2)}{\partial \overline{I}_2}\right),$$

$$\chi_2(\overline{I}_1,\overline{I}_2,\overline{I}_{1,\max}) = -2\frac{\partial^{\text{intact}}\Psi_{\text{ISO}}^{\infty}(\overline{I}_1,\overline{I}_2)}{\partial \overline{I}_2},$$

$$\xi_{1,\alpha}(t,T,\overline{I}_1,\overline{I}_2,\overline{I}_{1,\max}) = 2\left(\frac{\partial^{\text{intact}}\Upsilon_\alpha(t,T,\overline{I}_1,\overline{I}_2)}{\partial \overline{I}_1} + \overline{I}_1\frac{\partial^{\text{intact}}\Upsilon_\alpha(t,T,\overline{I}_1,\overline{I}_2)}{\partial \overline{I}_2}\right),$$

$$\xi_{2,\alpha}(t,T,\overline{I}_1,\overline{I}_2,\overline{I}_{1,\max}) = -\frac{\partial^{\text{intact}}\Upsilon_\alpha(t,T,\overline{I}_1,\overline{I}_2)}{\partial \overline{I}_2},$$

In this reformulated version of the generalized temperature- and Mullins damage-sensitive visco-hyperelastic constitutive framework, the total stress is rigorously constructed as a linear combination of tensorial integrity basis elements, each modulated by a scalar-valued response function. These functions encapsulate the combined effects of time evolution, thermal loading, deformation state, and internal degradation. Consequently, the individual components of stress are governed by the following coupled system of equations:

$$\left[\operatorname{vec}\left(\mathbf{S}_{\mathrm{VOL}}^{\infty}\right)\right]=\left[\operatorname{vec}\left(\mathbb{G}_{1}\right)\right]\left[\delta(\mathrm{J})\right] \tag{16}$$

$$\left[\operatorname{vec}\left(\frac{\mathbf{S}_{\mathrm{ISO}}^{\infty}}{J^{-2/3}(1-\vartheta)}\right)\right]=\left[\operatorname{vec}\left(\mathbb{G}_{2}\right)\ \operatorname{vec}\left(\mathbb{G}_{3}\right)\right]\begin{bmatrix}\chi_{1}(\overline{I}_{1},\overline{I}_{2},\overline{I}_{1,\max})\\ \chi_{2}(\overline{I}_{1},\overline{I}_{2},\overline{I}_{1,\max})\end{bmatrix}$$

$$\left[\operatorname{vec}\left(\frac{\sum_{\alpha=1}^{m}\mathbf{Q}_{\alpha}}{J^{-2/3}(1-\vartheta)}\right)\right]=\left[\operatorname{vec}\left(\mathbb{G}_{2}\right)\ \operatorname{vec}\left(\mathbb{G}_{3}\right)\right]\begin{bmatrix}\xi_{1,\alpha}(t,T,\overline{I}_{1},\overline{I}_{2},\overline{I}_{1,\max})\\ \xi_{2,\alpha}(t,T,\overline{I}_{1},\overline{I}_{2},\overline{I}_{1,\max})\end{bmatrix}$$

where vec(.) denotes the matrix Voigt form. To incorporate temperature dependence more efficiently and reduce the dimensionality of the response function inputs, the explicit dependencies on physical time t and temperature T in the response functions are now replaced by the reduced time variable $\tau^{*}$, as defined in the Time-Temperature Superposition principle (**Eq. 12**). This substitution relies on the assumption of thermo-rheological simplicity, allowing temperature effects to be captured through a unified time scale. In this reformulated framework, the isochoric equilibrium and non-equilibrium stress contributions are no longer treated separately. Instead, they are combined through a single generalized response function $\omega_{\alpha}(\tau^{*},\overline{I}_{1},\overline{I}_{2},\overline{I}_{1,\max})$, which compactly represents time-, temperature-, and history-dependent effects. This yields a more concise and temperature-normalized expression for the total isochoric stress contribution. The reformulated expression is:

$$\omega_{1,\alpha}(\tau^{*},\overline{I}_{1},\overline{I}_{2},\overline{I}_{1,\max})=\chi_{1}(\overline{I}_{1},\overline{I}_{2},\overline{I}_{1,\max})+\xi_{1,\alpha}(\tau^{*},\overline{I}_{1},\overline{I}_{2},\overline{I}_{1,\max}) \tag{17}$$
$$\omega_{2,\alpha}(\tau^{*},\overline{I}_{1},\overline{I}_{2},\overline{I}_{1,\max})=\chi_{2}(\overline{I}_{1},\overline{I}_{2},\overline{I}_{1,\max})+\xi_{2,\alpha}(\tau^{*},\overline{I}_{1},\overline{I}_{2},\overline{I}_{1,\max})$$

which is then used to construct the updated matrix form of stress components as:

$$\left[\operatorname{vec}\left(\mathbf{S}_{\mathrm{VOL}}^{\infty}\right)\right]=\left[\operatorname{vec}\left(\mathbb{G}_{1}\right)\right]\left[\delta(\mathrm{J})\right] \tag{18}$$

$$\left[\operatorname{vec}\left(\frac{\mathbf{S}_{\mathrm{ISO}}^{\infty}+\sum_{\alpha=1}^{m}\mathbf{Q}_{\alpha}}{J^{-2/3}(1-\vartheta)}\right)\right]=\left[\operatorname{vec}\left(\mathbb{G}_{2}\right)\ \operatorname{vec}\left(\mathbb{G}_{3}\right)\right]\begin{bmatrix}\omega_{1,\alpha}(\tau^{*},\overline{I}_{1},\overline{I}_{2},\overline{I}_{1,\max})\\ \omega_{2,\alpha}(\tau^{*},\overline{I}_{1},\overline{I}_{2},\overline{I}_{1,\max})\end{bmatrix}$$

The system above is structurally equivalent to **Eq.7** and is reformulated in the canonical linear form [b]=[A][x], where [x] encapsulates the unknown response functions as coefficient terms. This transformation enables the systematic identification of these functional coefficients via a least-squares optimization scheme ($min\|[A][x]-[b]\|_{2}^{2}$), facilitating a data-driven calibration of the constitutive model. The preceding formulation decomposes the total stress response of an isotropic, temperature- and history-dependent visco-hyperelastic material with Mullins effect into distinct physical contributions. The volumetric response, governed by δ(J), ensures thermodynamic consistency in compressibility. The isochoric stress contribution, which includes both equilibrium and non-equilibrium effects, is now represented through unified response functions $\omega_\alpha(\tau^*, \overline{I}_1, \overline{I}_2, \overline{I}_{1,\max})$. These functions compactly encode the combined influence of elastic and viscoelastic behavior, evolving over the reduced time, and account for deformation history, strain invariants, and Mullins effect. This structure enables the model to capture time-, temperature-, and damage-driven material evolution. By expressing the stress components through scalar-valued functions of invariants and internal variables, the formulation remains compact, interpretable, and compatible with ML frameworks. Unlike traditional models with predefined constitutive forms, this approach enables direct learning of response behavior from data, providing a flexible, physics-informed path for characterizing complex material phenomena.

## 3 Physics-Informed Constitutive Modeling Based on a Data-Driven Approach

### 3.1 Learning-Based Stress Mapping

Following **Eq.18**, the proposed physics-informed constitutive framework is structured around two ML surrogate models that represent the volumetric and isochoric stress contributions.

1. **Volumetric Surrogate Model** $\tilde{M}_{vol}$ (**Eq. 18$_a$**): It learns the bulk response function δ(J), which governs the compressive behavior of the material as a function of volume ratio *J*.

2. **Unified Isochoric Stress Surrogate Model** $\tilde{M}_{iso-visco-thermo-damage}$ (**Eq. 18$_b$**): Instead of treating equilibrium and non-equilibrium responses separately, this model learns a combined response function $\omega_\alpha(\tau^*, \overline{I}_1, \overline{I}_2, \overline{I}_{1,\max})$ that captures both elastic and viscoelastic stress contributions. The inputs are the reduced time $\tau^*$, the isochoric invariants $\overline{I}_1$ and $\overline{I}_2$, and the maximum value of the first invariant $\overline{I}_{1,\max}$. This maximum invariant is used to determine the maximum strain energy history, which plays a key role in distinguishing and capturing Mullins damage behavior. This unified representation enhances model compactness and reflects the influence of deformation history, temperature, and Mullins effect within a single learned mapping. These mappings are defined as:

$$\tilde{M}_{vol} : J \in \mathbb{R}^1 \rightarrow \delta \in \mathbb{R}^1 \tag{19}$$

$$\tilde{M}_{iso-visco-thermo-damage} : (\tau^*, I_1, I_2, \overline{I}_{1,\max}) \in \mathbb{R}^4 \rightarrow \vartheta, \sum \omega_{,\alpha} \in \mathbb{R}^3$$

Each surrogate model is trained on a dataset tailored to the physical mechanisms it represents:

- The volumetric model uses J as input and $\delta(J)$ as the training target to compute compressive stress.
- The isochoric model uses $\tau^*$, $\overline{I}_1$, $\overline{I}_2$, and $\overline{I}_{1,\max}$ as inputs to learn the full deviatoric stress response through the target variables of $\vartheta$, $\sum_{\alpha=1}^{m}\omega_{1,\alpha}$, and $\sum_{\alpha=1}^{m}\omega_{2,\alpha}$. This unified framework reduces redundancy, improves training efficiency, and maintains physical interpretability while capturing complex thermo-viscoelastic-damage interactions in soft materials.

Once the surrogate models are trained, the two constitutive stress components, volumetric and unified iso-visco-thermo-damage, can be evaluated as follows:

$$\tilde{\mathbf{S}}_{\text{VOL}}^{\infty,i} = \tilde{\delta}^i(\mathrm{J})\mathbb{G}_1^i \qquad (20)$$

$$\tilde{\mathbf{S}}_{\text{ISO}}^{\infty,j} + \sum_{\alpha=1}^{m}\tilde{\mathbf{Q}}_\alpha^k = J^{-2/3}(1-\vartheta)\left[\sum_{\alpha=1}^{m}\tilde{\omega}_{1,\alpha}^k(\tau^*,\overline{I}_1,\overline{I}_2,\overline{I}_{1,\max})\mathbb{G}_2^k + \sum_{\alpha=1}^{m}\tilde{\omega}_{2,\alpha}^k(\tau^*,\overline{I}_1,\overline{I}_2,\overline{I}_{1,\max})\mathbb{G}_3^k\right]$$

where the tilde symbol $\tilde{\bullet}$ indicates the scalar coefficients through prediction by the trained surrogate models.

## 3.2 Generation of Model Training Data

In practice, stress-strain data are typically acquired through experiments or high-fidelity simulations, offering direct insight into the material's macroscopic response. However, these measurements inherently entangle various stress contributions, volumetric, elastic, and rate-dependent effects, making it difficult to isolate the underlying mechanisms. To enable effective learning and physical interpretability, the raw mechanical response dataset $\mathcal{D}$ (containing stress, strain, and strain rate information) must be transformed into a structured dataset $\mathcal{D}^*$, which separates the contributions based on scalar invariants and surrogate model outputs and allows ML models to learn distinct, physically meaningful mappings. To align with the two surrogate models, i.e., the volumetric and unified isochoric visco-thermo-damage models, we construct $\mathcal{D}^*$ based on computed invariants and their corresponding response function targets. These invariants serve as the key input features for training and reflect both deformation history and damage evolution. Formally, the dataset is defined as:

$$\mathcal{D}^* \coloneqq \left\{J^i, \delta(J^i)\right\}_{i=1}^{N_{vol}} \cup \qquad (21)$$

$$\left\{\left[\tau^{*k},\overline{I}_1^k,\overline{I}_2^k,\overline{I}_{1,\max}^k\right],\left[\omega_{1,\alpha}(\tau^{*k},\overline{I}_1^k,\overline{I}_2^k,\overline{I}_{1,\max}^k),\omega_{2,\alpha}(\tau^{*k},\overline{I}_1^k,\overline{I}_2^k,\overline{I}_{1,\max}^k)\right]\right\}_{k=1}^{N_{iso-visco-thermo-damage}}$$

Here, the first term corresponds to the volumetric surrogate model and the second captures the unified isochoric component, including rate- and damage-dependent effects through the learned response functions $\omega_\alpha(\tau^*, \bar{I}_1, \bar{I}_2, \bar{I}_{1,\max})$. Given that the model is grounded in stress decomposition, it is critical to ensure consistency between the physical phenomena represented in real-world data and the structure of the training datasets. This becomes particularly important in the presence of history-dependent behaviors like viscoelastic relaxation or damage progression, which significantly affect the long-term material response. The following subsection presents a systematic methodology for constructing such consistent and physically meaningful training datasets.

### 3.2.1 Approach for Total Stress Decomposition

To prepare the training dataset for the proposed two-part surrogate modeling framework, we adopt a stress decomposition strategy based on physically meaningful measures. The total stress recorded from experiments or simulations includes both volumetric and isochoric contributions, with the latter encompassing time-dependent viscoelastic effects as well as the equilibrium response.

To isolate the volumetric stress component, a hydrostatic compression test can be performed, where the material is compressed uniformly in all directions. Because this loading path induces no shear, the resulting stress is purely volumetric, allowing us to directly obtain $\mathbf{S}_{\text{VOL}}^\infty$.

The remaining part of the stress, obtained by subtracting the volumetric component from the total stress, is attributed to the unified isochoric response, which includes both the equilibrium and time-dependent contributions. Formally, $\mathbf{S}_{\text{ISO}} = \mathbf{S}_{\text{total}} - \mathbf{S}_{\text{VOL}}$. This unified stress term is used to train a single surrogate model that captures complete isochoric behavior, including effects due to viscoelasticity, temperature via reduced time, and Mullins damage. This decomposition is repeated for all loading paths and time instances to construct a comprehensive dataset. The resulting volumetric and isochoric stress components are then used along with the computed integrity basis tensors to solve the linear system in **Eq. 18**. The extracted response functions combined with scalar invariants form the dataset $\mathcal{D}^*$, which serves as the training data for the surrogate models introduced in **Eq. 19**.

## 3.3 Embedding Physics Principles in the Data-Driven Constitutive Framework

Due to the structured formulation of the proposed data-driven constitutive framework (**Eq. 19**) and its stress computation strategy (**Eq.20**), several core physical principles are inherently enforced:

- ❖ **Material Frame Indifference (Objectivity)**

Strain energy is unaffected by superimposed rigid body rotations, ensuring objectivity $\Psi(\mathbf{C}) = \Psi(\mathbf{QCQ}^T), \forall \mathbf{Q} \in SO(3)$. The model satisfies this by using $\mathbf{C}$, making the integrity basis in **Eq.20** objective.

❖ **Material Symmetry**

Isotropy is preserved through the use of principal invariants $I_1$, $I_2$, and $J$ (**Eq.19**). $\Psi(\mathbf{C})$ remain invariant under all symmetric rotations.

$$\Psi(\mathbf{C}) = \mathbf{Q}^T \Psi(\mathbf{C}) \mathbf{Q}, \forall \mathbf{Q} \in SO(3).$$

❖ **Angular Momentum Balance**

Because the integrity basis tensors in **Eq.20** are symmetric by design, the stress tensor $\mathbf{S}$ remains symmetric as required ($\mathbf{S}=\mathbf{S}^T$). This symmetry upholds the consistency of both second Piola-Kirchhoff and Cauchy stresses, ensuring no artificial internal moments are introduced.

In addition to these embedded constraints, three further conditions are explicitly imposed in the ML surrogate models:

I. **Stress-Free Reference Configuration**: At $\mathbf{C}=\mathbf{I}$, the equilibrium stress must vanish: $\mathbf{S}_{\text{ISO}}^{\infty} = \mathbf{S}_{\text{VOL}} = \mathbf{0}$. This is imposed within the $\tilde{\mathrm{M}}_{vol}$ surrogate model.

II. **Thermodynamic Consistency**: The viscoelastic dissipation must be non-negative (**Eq. 7e**), enforced in the TCN-based model by constraining the internal variables accordingly in $\tilde{\mathrm{M}}_{iso-visco-thermo-damage}$.

III. **Damage Boundedness**: The damage parameter, representing material integrity, is constrained to the interval $0 \leq \vartheta \leq 1$, ensuring a physically reasonable Mullins damage.

## 3.4 Modeling Equilibrium Volumetric Response ($\tilde{\mathrm{M}}_{vol}$)

To model the equilibrium volumetric stress component, several approaches can be used with a built-in stress-free reference condition $\mathbf{S}_{\text{VOL}}(\mathbf{C}=\mathbf{I})=\mathbf{0}$. The mappings for volumetric stress δ(J) and isochoric stress coefficients with normalization can be applied to ensure that all stress components vanish in the undeformed configuration (J=1 and δ(J)=0). Several studies have employed Gaussian Process Regression (GPR) to model the volumetric response of materials. GPR is utilized to capture the volumetric stress behavior, see [144] for further methodological details. In the current work, although the proposed model is capable of representing volumetric effects, we intentionally exclude them and focus on incompressible polymers to avoid redundancy and isolate the isochoric response.

### 3.5 Modeling History-Dependent Material Response ($\tilde{M}_{iso-visco-thermo-damage}$) via TCN

Temporal Convolutional Networks or TCNs provide efficient and scalable architecture for modeling sequential data, particularly in settings where preserving causal structure is critical. Unlike traditional recurrent networks, TCNs can process sequences of arbitrary length while maintaining a consistent input-output sequence size [98], [104]. TCNs use causal convolutions, meaning the output $\hat{y}_t$ (here Stress) at time $t$ depends only on the current and past inputs $x_t$ (here $\tau, I_1, I_2, I_{1,\max}$), not on future ones. This prevents information from leaking backward in time. In simple terms, the model learns a mapping:

$$\hat{y}_0, ..., \hat{y}_t = f(x_0, ..., x_t) \tag{22}$$

The architecture used in TCN is a 1D fully convolutional network (FCN). In the TCN structure, to achieve long-range temporal dependencies without increasing network depth excessively, dilated convolutions are introduced. For a 1D input $x \in \mathbb{R}^n$ and filter $f : \{0, \ldots, k-1\} \to \mathbb{R}$, the dilated convolution is defined as:

$$F(s) = (\mathbf{x}_{*_d} f)(s) = \sum_{i=0}^{k-1} f(i) \cdot \mathbf{x}_{s-d \cdot i} \tag{23}$$

where $k$ is the filter size and *$d$ means a dilated convolution, with dilation factor $d$ controlling the spacing between filter taps. By increasing *$d$ exponentially across layers, TCNs effectively extend their receptive field, allowing them to capture long-range patterns even with relatively shallow architectures [104]. As shown in **Figure 2(a)**, dilation introduces gaps between filter taps, enabling each convolutional layer to cover a wider range of past inputs. To improve training stability, TCNs employ residual connections. A residual block applies a transformation $\mathbb{N}$ to the input and adds it back:

$$O = Activation(\mathbf{x} + \mathbb{N}(\mathbf{x})) \tag{24}$$

In **Figure 2b** each residual block in a TCN processes the input sequence using two layers of dilated causal convolutions, each followed by weight normalization, ReLU activation, and dropout to capture long-range dependencies while ensuring stability and regularization. A 1×1 convolution is applied in parallel to match dimensions, and the two paths are added together to enhance learning and enable deeper networks. Each residual block refines the internal sequence representation, which encodes the history-dependent behavior of the material. To produce the final output for each time step, an FCN is employed as a decoder, mapping the encoded sequence to target variables such as internal state components or constitutive quantities (**Figure 2c**). This separation between temporal feature extraction (encoder) and output mapping (decoder) provides flexibility in modeling history-dependent responses. Additionally, the loss function incorporates dissipation-based penalties to enforce physical constraints. Compared to conventional architectures, the TCN design used here avoids gated units and skip connections, resulting in a simpler, efficient, yet expressive structure.

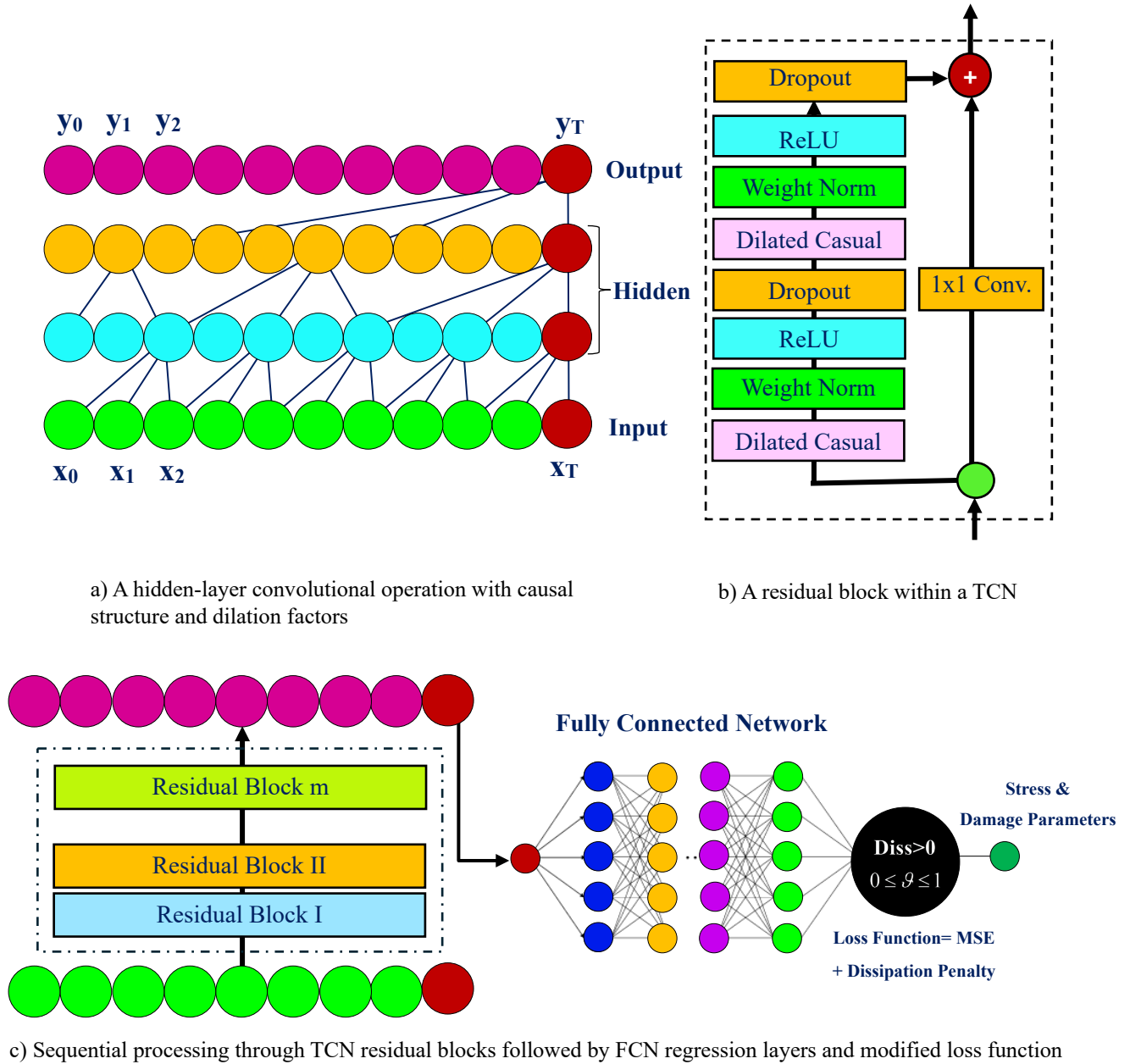

a) A hidden-layer convolutional operation with causal structure and dilation factors

b) A residual block within a TCN

c) Sequential processing through TCN residual blocks followed by FCN regression layers and modified loss function

Figure 2: Architecture of TCN.

### 3.5.1 Construction of the loss function

The TCN-based framework for modeling path-dependent material behavior incorporates the Clausius-Duhem inequality (**Eq.7e**) as a hard thermodynamic constraint to ensure physical consistency. This inequality, central to the second law of thermodynamics, governs irreversible processes by restricting the admissible evolution of internal variables and ensuring non-negative energy dissipation. Incorporating this constraint into the TCN architecture anchors the model in thermodynamic law, ensuring that predictions involving temperature effects and damage evolution remain physically plausible. This enables the model to go beyond standard invariance requirements and capture complex, history-dependent behavior without violating energetic principles. In the internal variable-based viscoelastic framework, the evolution of material behavior away from equilibrium is captured through a configurational free energy function $^{\text{intact}}\Upsilon_\alpha(\overline{\mathbf{C}},\boldsymbol{\Gamma}_\alpha)$. The thermodynamic driving force for each non-equilibrium branch is derived from this energy potential and can be expressed in two equivalent forms: as a derivative with respect to $\overline{\mathbf{C}}$, and as a negative derivative with respect to $\boldsymbol{\Gamma}_\alpha$. This reflects the conjugate relationship between stress and internal variables and ensures compliance with the Clausius-Duhem inequality [145].

$$\mathbf{Q}_\alpha = 2\frac{\partial\, ^{\text{intact}}\Upsilon_\alpha(\overline{\mathbf{C}},\boldsymbol{\Gamma}_\alpha)}{\partial \overline{\mathbf{C}}} = -2\frac{\partial\, ^{\text{intact}}\Upsilon_\alpha(\overline{\mathbf{C}},\boldsymbol{\Gamma}_\alpha)}{\partial \boldsymbol{\Gamma}_\alpha} \tag{25}$$

The resulting expression establishes a consistent thermodynamic link between the stored energy, deformation, and internal evolution, thus, regarding the above expression, **Eq.7e** yields:

$$\mathrm{D}_{\text{int}} = \dot{\vartheta}\left(^{\text{intact}}\Psi_{\text{ISO}}^\infty(\overline{\mathbf{C}}) + \sum_{\alpha=1}^{m}\, ^{\text{intact}}\Upsilon_\alpha(\overline{\mathbf{C}},\boldsymbol{\Gamma}_\alpha)\right) + \frac{(1-\vartheta)}{2}\sum_{\alpha=1}^{m}\mathbf{Q}_\alpha:\dot{\boldsymbol{\Gamma}}_\alpha \geq 0 \tag{26}$$

Following the approach outlined by Holzapfel [145], the configurational free energy function and internal deviatoric history variables are defined in a specific form to describe the time-dependent, isochoric contribution of each non-equilibrium branch, considering the damage variable in the strain energy function, which reaches

$$\sum \Upsilon_\alpha(\overline{\mathbf{C}},\boldsymbol{\Gamma}_\alpha) = \sum\left[\beta_\alpha^\infty\, ^{\text{intact}}\Psi^\infty(\overline{\mathbf{C}}) - \mu_\alpha|\boldsymbol{\Gamma}_\alpha|^2\right], \tag{27}$$

$$\boldsymbol{\Gamma}_\alpha = \frac{(1-\vartheta)}{\mu_\alpha}J^{-2/3}\mathrm{Dev}\left(\nabla_{\overline{\mathbf{C}}}\, ^{\text{intact}}\Psi^\alpha(\overline{\mathbf{C}})\right) - \frac{1}{2\mu_\alpha}\mathbf{Q}_\alpha$$

where $\mu_\alpha$ represents a non-negative, temperature-dependent parameter, which is fixed to unity in this study to isolate the structural behavior of the model. While the data-driven framework does not rely on a predefined constitutive law, this specific expression is deliberately embedded, drawing directly from the thermodynamically consistent structure of the Holzapfel model [28], to rigorously enforce the second law of thermodynamics. The formulation establishes a precise balance between the stored energy and internal dissipation: the first term governs the isochoric stress response, while the second term imposes a strict dissipative constraint. This ensures that the evolution of internal variables remains energetically admissible, stabilizing the model and preserving its physical fidelity under all admissible loading conditions.

Considering the isochoric equilibrium stress defined as $\mathbf{S}_{iso}^{\infty} = J^{-2/3}(1-\vartheta)\text{Dev}\left(2\nabla_{\overline{\mathbf{C}}}{}^{\text{intact}}\Psi^{\infty}(\overline{\mathbf{C}})\right)$, and the specific form of the internal isochoric stress component $\mathbf{S}_{iso,\alpha} = \beta_{\alpha}^{\infty}\mathbf{S}_{iso}^{\infty}$ where $\beta_{\alpha}^{\infty}$ is a scaling factor associated with the energy potential; we follow the structure proposed in the Holzapfel model [28] using **Eq. 27b**, along with the derivative of the internal variable $\Gamma_{\alpha}$ to obtain:

$$\dot{\Gamma}_{\alpha} = \frac{1}{2}\left(\beta_{\alpha}^{\infty}\dot{\mathbf{S}}_{iso}^{\infty} - \dot{\mathbf{Q}}_{\alpha}\right) \tag{28}$$

In computational implementation, the time derivative of the internal variable at a discrete time step is typically approximated using a finite difference method by a time integration scheme, such as the backward Euler method; thus, we have

$$\dot{\Gamma}_{\alpha} = \frac{1}{2\Delta t}\left[\beta_{\alpha}^{\infty}\left(\mathbf{S}_{iso}^{\infty}\big|_{t_{n+1}} - \mathbf{S}_{iso}^{\infty}\big|_{t_n}\right) - \left(\mathbf{Q}_{\alpha}\big|_{t_{n+1}} - \mathbf{Q}_{\alpha}\big|_{t_n}\right)\right] \tag{31}$$

TCN directly predicts the equilibrium isochoric stress, which is used in **Eq.26** to compute viscous dissipation at each time step. To ensure thermodynamic consistency, the loss function penalizes any violation of the non-negativity condition on this dissipation. **Figure 3** presents the physics-informed training framework, where the loss function combines standard prediction error with penalties that enforce physical constraints. Specifically, non-negativity of viscous dissipation and boundedness of the damage variable are both imposed; violations of either condition introduce penalty terms that steer TCN weight updates during optimization.

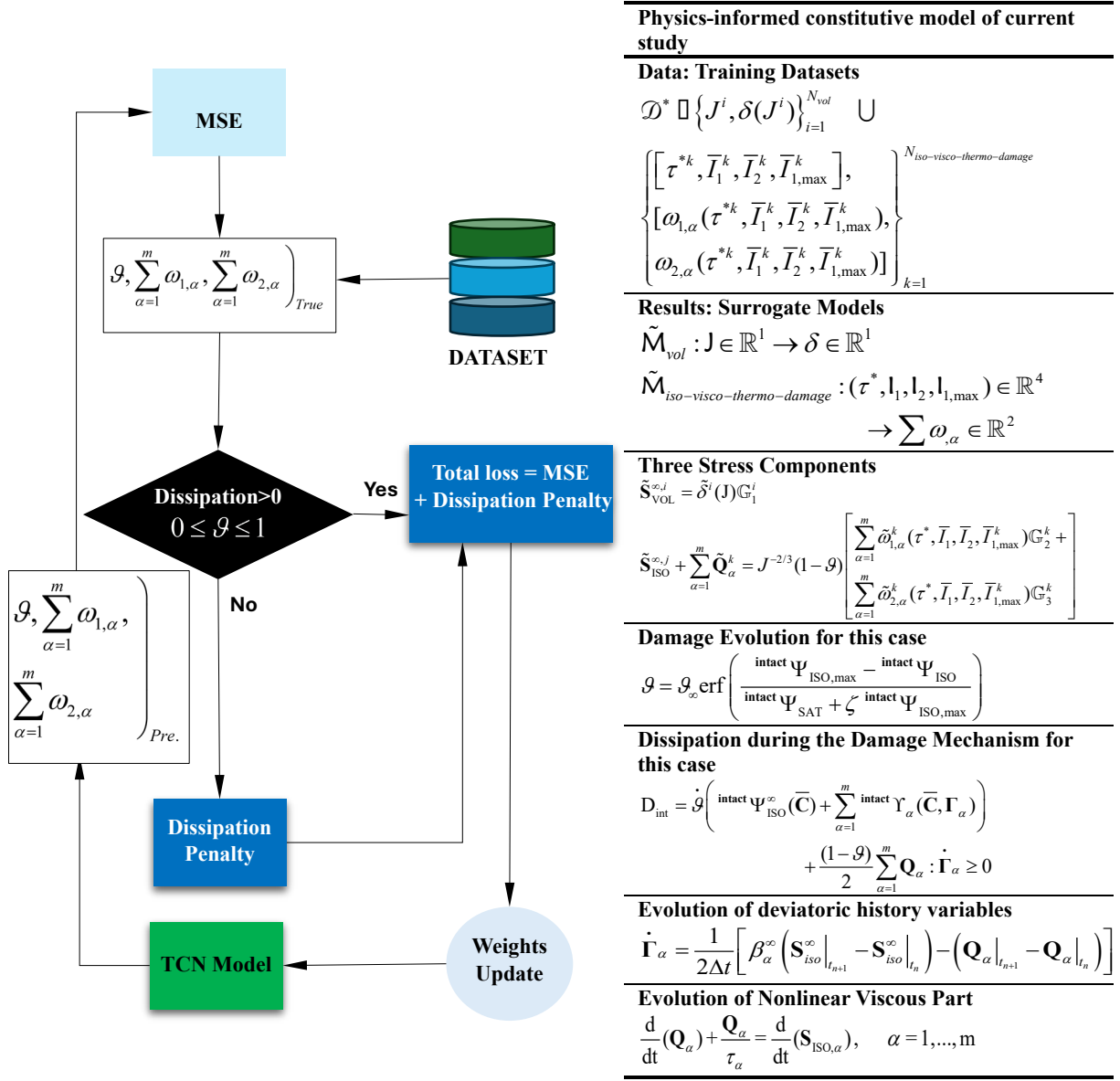

Figure 3. Workflow of the TCN model incorporating a physics-informed loss function that combines data-driven prediction error with dissipation and damage constraints

## 4 Assessment of Surrogate Model Performance

### 4.1 Dataset Generation for Model Training

The framework employs a physics-informed surrogate model including a TCN that learns the time-, temperature-, and damage-dependent viscoelastic and isochoric response. Unlike prior methods that decouple deformation modes, our approach preserves their natural coupling by training on data collected across multiple time scales and temperatures. The dataset is grounded in constitutive formulations that incorporate thermal effects, damage evolution, and energy consistency.

### 4.1.1 Equilibrium part

We examine the material response under uniaxial loading, where the principal stretch $\lambda_1$ is applied, while the Yeoh model $\Psi_{ISO}^{\infty}(\overline{\mathbf{C}}) = C_{10}(\overline{I}_1 - 3) + C_{20}(\overline{I}_1 - 3)^2 + C_{30}(\overline{I}_1 - 3)^3$, governs the isochoric (deviatoric) stress response.

### 4.1.2 Softening Damage part

Ogden and Roxburgh [146] proposed a formulation to embed the Mullins effect into hyperelastic models by introducing an internal scalar variable that tracks material degradation. This damage variable, $\vartheta \in [0,1]$, evolves based on the material's deformation history and represents isotropic softening. The damage is linked to the isochoric strain energy and evolves according to:

$$\vartheta = \vartheta_{\infty} \operatorname{erf}\left(\frac{{}^{\text{intact}}\Psi_{ISO,\max} - {}^{\text{intact}}\Psi_{ISO}}{{}^{\text{intact}}\Psi_{SAT} + \zeta\, {}^{\text{intact}}\Psi_{ISO,\max}}\right) \tag{32}$$

The related parameters are introduced in section 2.2. On unloading and reloading, unless the applied stretch exceeds its prior peak $\lambda_{max}$ the damage variable remains unchanged. Once the stretch exceeds $\lambda_{max}$, the maximum energy is updated, and $\vartheta$ increases accordingly. This evolution follows a piecewise, on-off behavior, with abrupt transitions triggered by new loading maxima.

### 4.1.3 Non-Equilibrium part

To capture nonlinear viscoelastic behavior under large strains and high stress levels, the Holzapfel model formulates evolution equations that describe both energy dissipation and time-dependent behavior. In this setting, each viscoelastic branch $\alpha$ includes a deviatoric internal variable $\mathbf{Q}_\alpha$ representing the non-equilibrium stress contributions associated with the internal history variable $\Gamma_\alpha$. The dissipative response is governed by the following first-order evolution law:

$$\frac{d}{dt}(\mathbf{Q}_\alpha) + \frac{\mathbf{Q}_\alpha}{\tau_\alpha} = \frac{d}{dt}(\mathbf{S}_{ISO,\alpha}), \quad \alpha = 1,...,m \tag{33}$$

This equation provides the foundation for generating synthetic training data representing non-equilibrium stress evolution. The viscoelastic network is modeled using $m$ parallel chains, each with its own relaxation time $\tau_\alpha \in (0,\infty)$, capturing a spectrum of time-dependent effects. Since soft materials consist of structurally similar chain networks, the isochoric strain energy for each branch is scaled from the equilibrium response using a deformation-dependent energy factor $\beta_{\infty}^{\alpha} \in (0,\infty)$, leading to

$^{\text{intact}}\Psi_{\text{ISO}}^{\alpha}(\overline{\mathbf{C}}) = \beta_{\infty}^{\alpha}\ ^{\text{intact}}\Psi_{\text{ISO}}^{\infty}(\overline{\mathbf{C}})$. This scaling ensures that each branch contributes proportionally to the total isochoric energy based on its own relaxation behavior. An explicit analytical solution to **Eq. 33** for the non-equilibrium internal variable is given by:

$$\mathbf{Q}_\alpha = \exp(-s/\tau_\alpha)\mathbf{Q}_\alpha^{0+} + \int_{0^+}^{t=s} \exp\left[-(s-t)/\tau_\alpha\right]\dot{\mathbf{S}}_{\text{ISO},\alpha}\, dt, \qquad (34)$$

with the initial condition $\mathbf{Q}_\alpha\big|_{t=0} = 0$. While this solution captures the exact time evolution, its direct application in numerical simulations is limited. For practical implementation, it must be reformulated into a robust incremental scheme that updates internal quantities over discrete time steps. Assuming $t_{n+1} = \Delta t + t_n$, the equation can be discretized accordingly to enable recursive evaluation in a computational framework.

$$\mathbf{S} = J_{n+1}\frac{\partial \Psi_{\text{VOL}}^{\infty}(J_{n+1})}{\partial J_{n+1}}\mathbf{C}_{n+1}^{-1} + 2\frac{\partial^0\Psi_{\text{ISO}}^{\infty}(\overline{\mathbf{C}}_{n+1}^{-1})}{\partial \mathbf{C}_{n+1}} + \sum_{\alpha=1}^{m}\mathbf{Q}_{\alpha,n+1} \qquad (35)$$

$$\text{where } \mathbf{Q}_{\alpha,n+1} = \int_{0^+}^{t=s_n}\exp\left[-(s_n-t)/\tau_\alpha\right]\dot{\mathbf{S}}_{\text{ISO},\alpha}\,dt + \int_{t=s_n}^{t=s_{n+1}}\exp\left[-(s_{n+1}-t)/\tau_\alpha\right]\dot{\mathbf{S}}_{\text{ISO},\alpha}\,dt$$

Applying the Mean Value Theorem to the separated convolution integral leads to the following simplified expression:

$$\mathbf{Q}_{\alpha,n+1} = \exp\left(-\frac{\Delta t}{\tau_\alpha}\right)\mathbf{Q}_{\alpha,n} + \exp\left(-\frac{\Delta t}{2\tau_\alpha}\right)\beta_\infty^{\alpha}\left(\mathbf{S}_{\text{ISO},n+1}^{\infty} - \mathbf{S}_{\text{ISO},n}^{\infty}\right) \qquad (36)$$

To focus on the key aspects of damage evolution and temperature dependence, we restrict our study to a comprehensive dataset. This dataset is generated using a temperature-dependent viscoelastic model incorporating stress-softening damage and assuming material incompressibility. It is constructed based on a four-branch generalized Maxwell model to capture the multi-timescale behavior typical of soft polymers. Each branch is defined with a distinct relaxation time and weighting factor to simulate both short-term dissipation and long-term memory effects. By embedding this dataset into a physics-informed TCN framework, we aim to evaluate the model's ability to capture history-dependent stress response, internal variable evolution, and satisfying dissipation behavior under thermomechanical loading.

### 4.2 TCN Architecture Design

The ML framework employs a physics-informed TCN to model the evolution of internal variables under thermomechanical loading. The TCN architecture consists of three dilated causal convolutional layers with 128, 64, and 32 filters, respectively, each with residual connections and dropout regularization. Input and target data are normalized using min-max scaling and structured into fixed-length temporal sequences.

A physics-informed loss function is employed during training, which not only minimizes the prediction errors for the internal variables but also accounts for dissipation consistency (**Figure 2c**). The model is trained using the Adam optimizer over 100 epochs with a batch size of 64. Results show that using only three tests is sufficient to accurately predict the full cyclic behavior of the material across varying stretches, temperatures, and strain rates. The model demonstrates high prediction accuracy, consistent dissipation behavior, and strong robustness across multiple thermal loading cycles.

## 4.3 Model Performance and Several Case Analyses

We examine the material behavior under cyclic uniaxial loading, where the principal stretch evolves independently and the transverse stretches satisfy incompressibility. To assess the accuracy of the surrogate model in each scenario, we employ the percent relative error metric defined as

$$\left( \left\| (vec(\mathbf{S}) - vec(\tilde{\mathbf{S}})) \right\| \Big/ \left\| vec(\tilde{\mathbf{S}}) \right\|_F \right) \times 100 \tag{24}$$

where $\|\cdot\|_F$ denotes the Frobenius norm. Here, $\tilde{\mathbf{S}}$ represents the predicted stress tensor from the surrogate model, and $\mathbf{S}$ is the corresponding true stress tensor. Pointwise errors are measured to identify the degree of mismatch, and their mean provides a cumulative indicator of the model's deviation across all cases $ERR = \frac{\sum_{i=1}^{N} ERR_i}{N}$, where N is the number of data samples used in either the training or testing phase.

We have considered different cases to comprehensively evaluate the surrogate model in the context of thermomechanical modeling, damage evolution, and generalization, including:

- *Model Training and Evaluation on Purely Unseen Stretch Levels*
- *General Evaluation Across Unseen Stretch, Temperature, and Strain Rate*
- *Thermodynamic Consistency and Noise Robustness of the Surrogate Model*
- *Damage Evolution in an Open-Hole Sample under High-Temperature Cyclic Loading*

### 4.3.1 Case 1: Model Training and Evaluation on Purely Unseen Stretch Levels

Data is obtained at three different temperatures (0°C, 25°C, and 55°C) and three strain rates (20, 200, and 500 %/min). These data are employed to investigate model performance, damage progression, and physical consistency. Each test included three loading–unloading cycles with very large deformations (100%, 150%, and 200% strain) to capture stress-softening damage behavior.

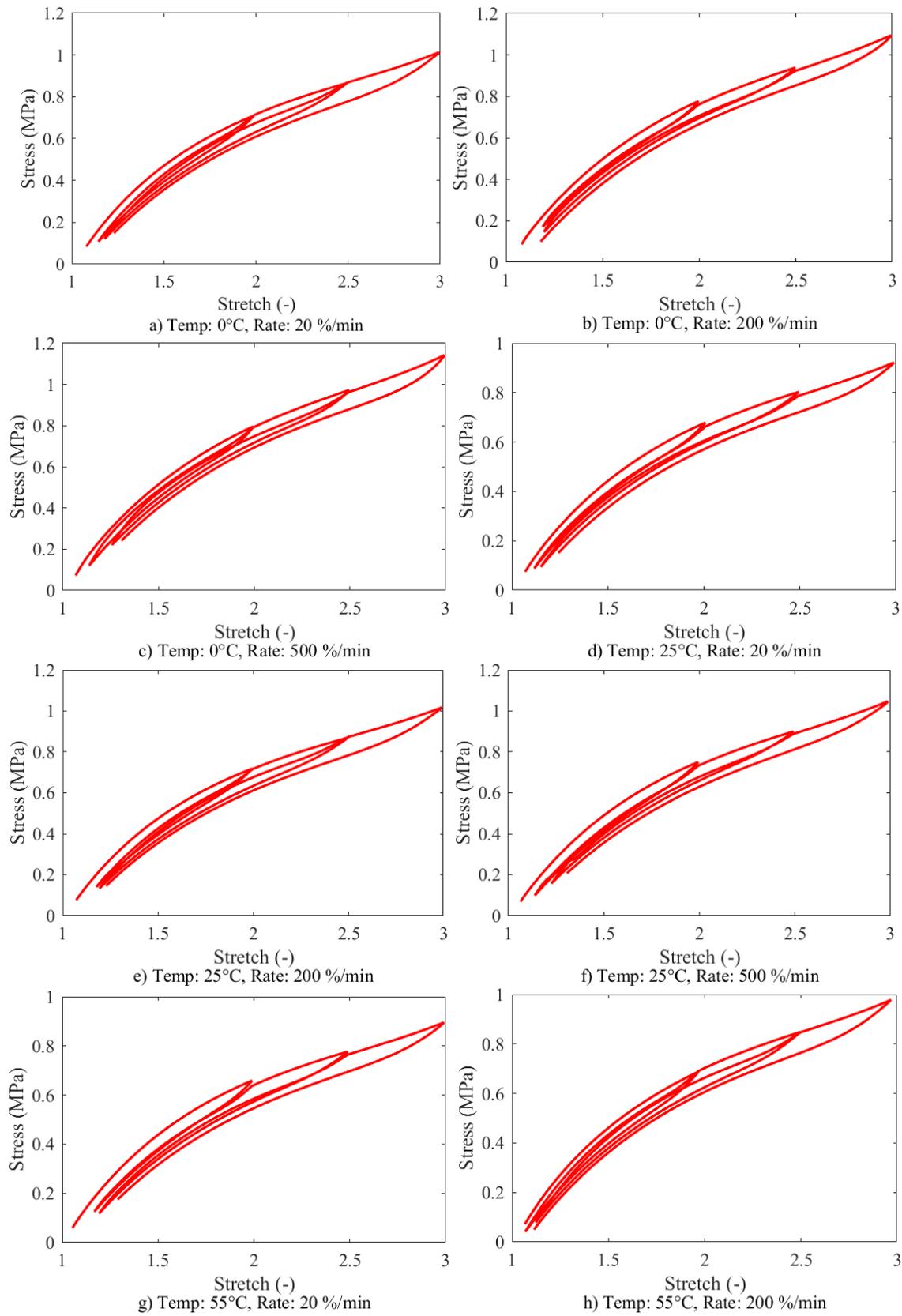

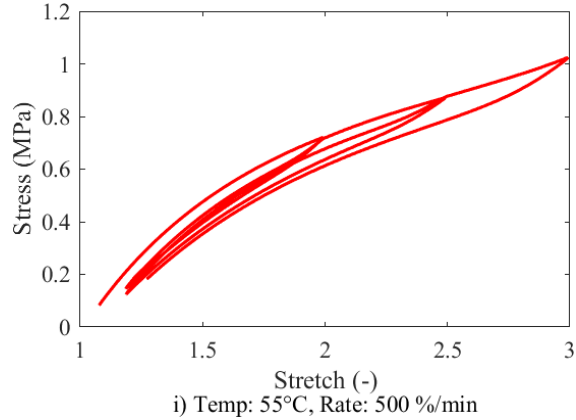

i) Temp: 55°C, Rate: 500 %/min

Figure 4: Mullins damage in shape memory polymers for different times and temperatures

For training, only three of the tests were utilized, and from each, only the first two loading–unloading cycles were included, just enough to inform the surrogate model of the material's viscoelastic behavior and the onset of damage evolution. Specifically, we selected three tests conducted under distinct thermomechanical conditions: (1) 0°C at 200%/min, (2) 25°C at 20%/min, and (3) 55°C at 200%/min, each loaded up to 150% strain. This selection was intentionally limited to span a diverse but minimal subset of the available data, ensuring that the model could learn from essential trends without overfitting to specific cases. The third cycle from each test was excluded from training and instead reserved to evaluate the model's generalization capability on unseen stretches, as stress-softening damage tends to manifest more clearly in successive cycles. This strategic separation of training and evaluation phases enables a focused examination of the model's ability to capture cumulative damage, path-dependent behavior, and the nonlinear response under increasing deformation levels across varying temperatures and strain rates.

As illustrated in **Figure 5**, the model demonstrates excellent agreement between predicted and true values for the response function $\omega_1$, $\omega_2$, and the stress-softening damage coefficient. The surrogate model was trained for 100 epochs, during which the loss function decreased rapidly within the first 20 epochs and gradually stabilized. The loss plateau observed beyond 60 epochs suggests that the model had already captured the essential physics governing viscoelastic and damage responses. Additionally, the convergence of the loss function during training confirms the surrogate model's efficiency and highlights its strong potential, even with a limited dataset and relatively short training duration.

Comparison between the ground truth and the predictions of the trained TCN-based surrogate model for two representative training cases: (e) 0°C at 200%/min and (f) 55°C at 200%/min. The surrogate model demonstrates excellent agreement with the stress response during both cycles, with high $R^2$ values of 0.9559 and 0.9502, respectively. The model successfully captures viscoelastic hysteresis, peak stress, and recovery behavior across cycles. Additionally, the decomposed stress contributions from the first and second deviatoric Maxwell branches are shown, confirming the model's ability to replicate the multi-branch relaxation behavior.

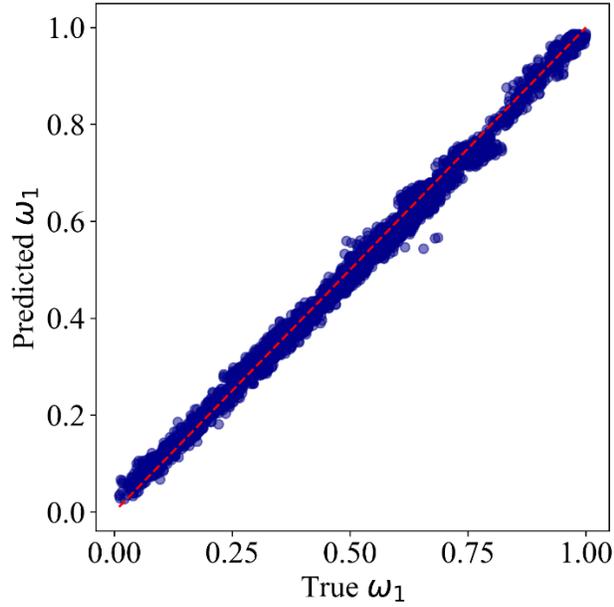

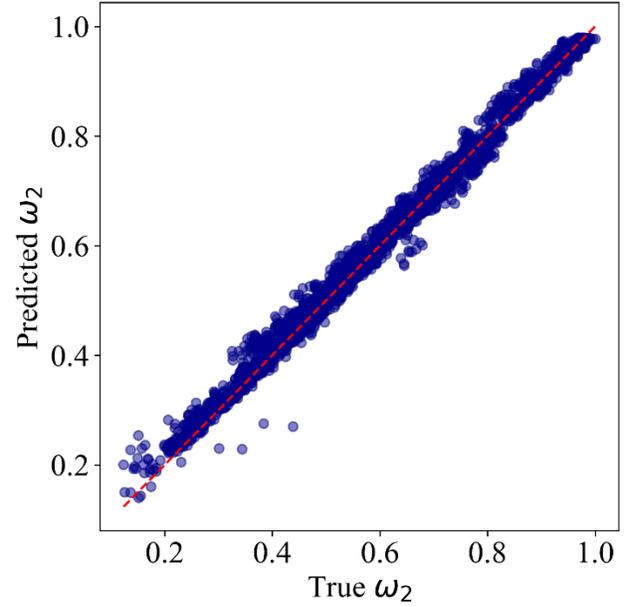

a) Predicted vs. true values of the first response function

b) Predicted vs. true values of the second response function

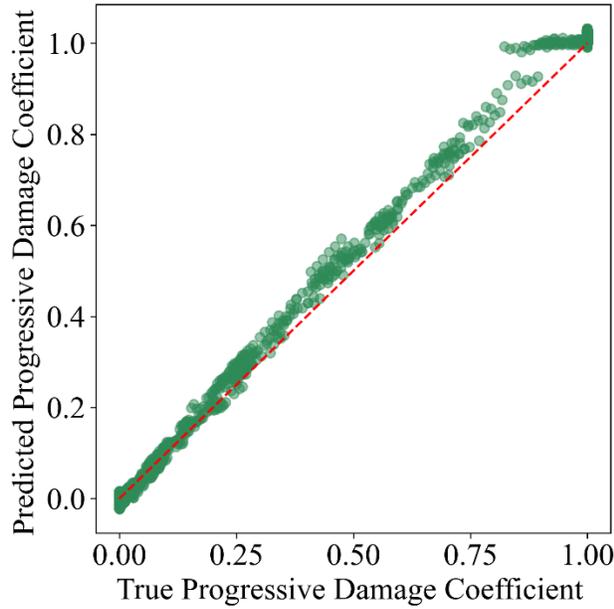

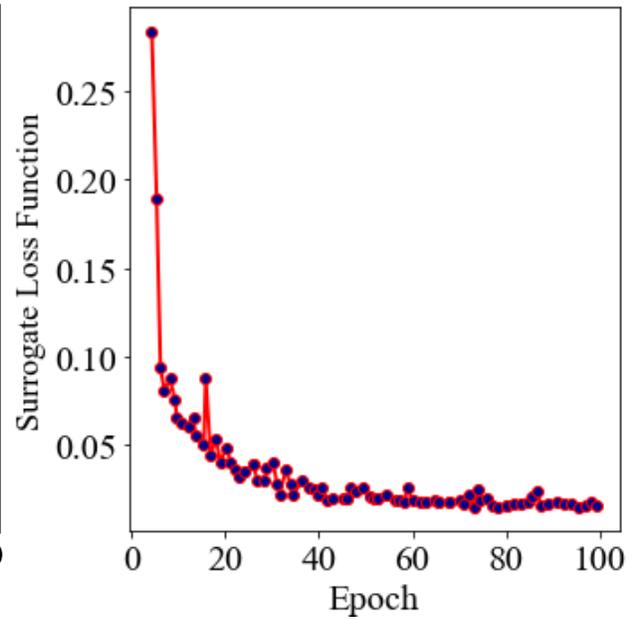

c) Predicted vs. true values of the damage coefficient

d) Evolution of the loss function over training epochs

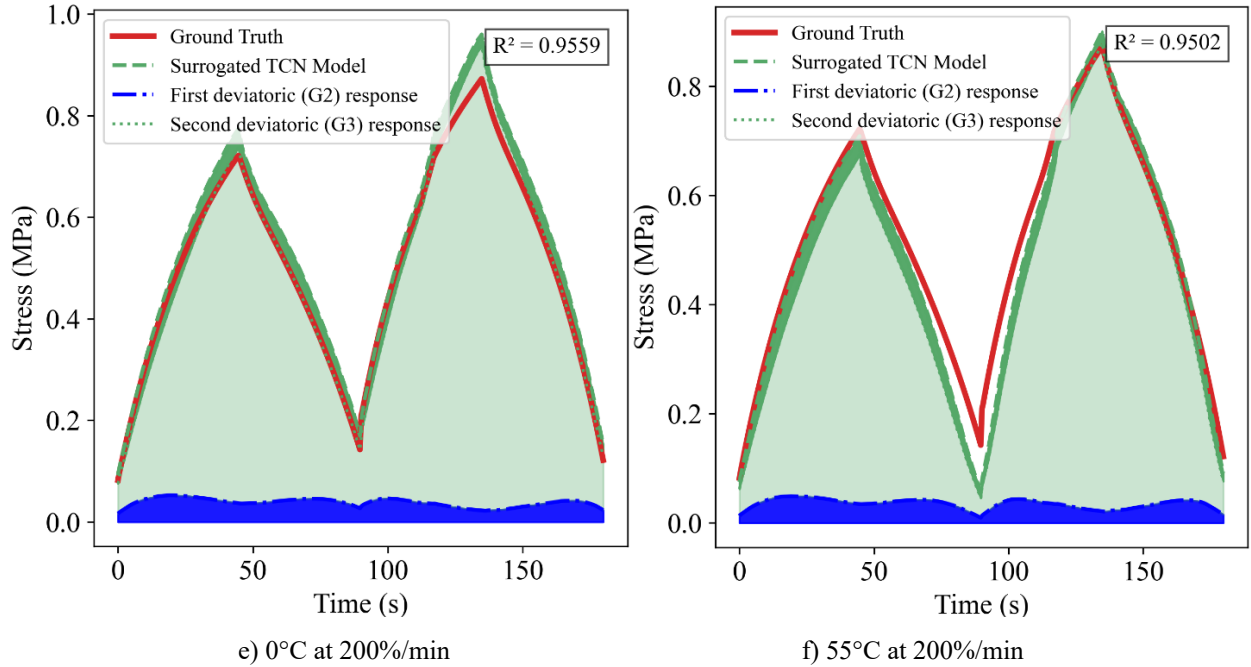

e) 0°C at 200%/min    f) 55°C at 200%/min

Figure 5: Surrogate model accuracy in predicting material parameters, damage, loss convergence, and stress decomposition under varying thermomechanical conditions

**Figure 6 (a)** and **(b)** show stretch–stress responses at 25°C and 20%/min and 55°C and 200%/min, respectively, while subfigures **(c)** and **(d)** present the corresponding stress–time responses at 0°C and 200%/min and 25°C and 20%/min. In each case, despite the third loading–unloading cycle reaching a stretch level nearly triple that of the initial cycle, the surrogate model accurately predicts the nonlinear material behavior, including loading–unloading hysteresis, viscoelastic recovery, and stress-softening damage accumulation. Moreover, the decomposed contributions from the first and second deviatoric branches are shown in the stress–time plots, reflecting the model's ability to capture internal viscoelastic mechanisms. These results emphasize the surrogate model's capability to provide physically meaningful predictions under large and previously unseen loading paths.

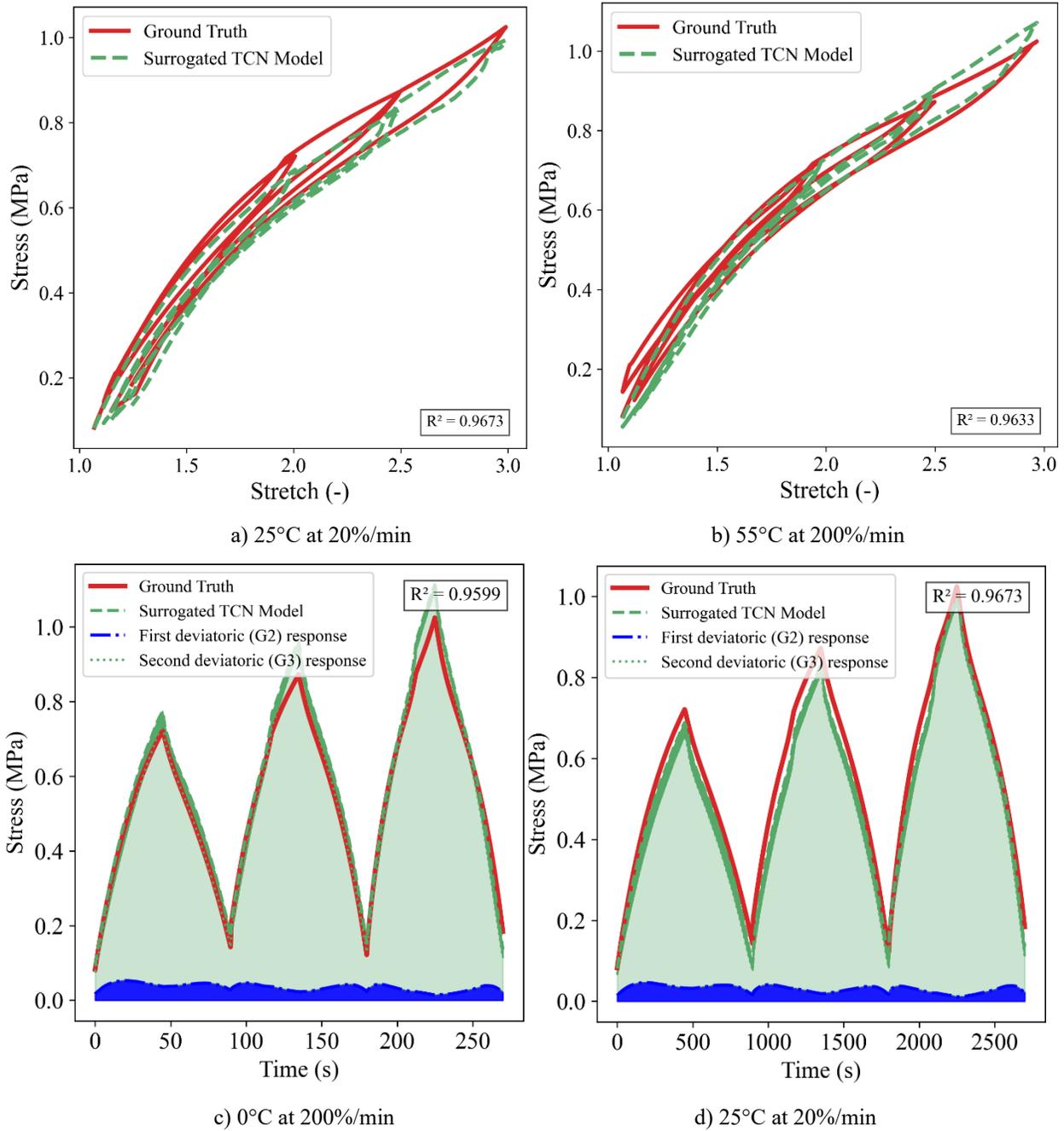

Figure 6: Comparison of predicted vs. ground truth stress and component decomposition across different temperatures and strain rates.

### 4.3.2 Case 2: Generalization to Unseen Conditions

In this case, the model's generalization capability is evaluated using unseen thermomechanical loadings. To evaluate the performance of the proposed surrogate model, we tested it against several additional loading conditions and one entirely new test at 10°C–500%/min, none of which were used during training. These test conditions spanned a broad range of thermo-mechanical states, including:

- Varying temperatures: 0°C, 10°C, 25°C, and 55°C
- Multiple strain rates: 20%, 200%, and 500% per minute
- Increasing stretch levels approaching 3.0

**Figure 7 (a)–(f)** displays the stress versus stretch responses for these unseen tests. Across all conditions, the surrogate TCN model closely tracks the ground truth curves, including the nonlinear loading–unloading hysteresis, with R² values consistently exceeding 0.97. These results indicate the model's capability to extrapolate over both strain rate and temperature domains beyond its training range. Even at low temperatures (0°C and 10°C) where viscoelastic time constants are smaller, the model maintains high prediction accuracy, demonstrating that the surrogate correctly encodes temperature-dependent relaxation behavior.

Furthermore, the stress–time plots in the second set of **Figure 8** provide deeper insight into the model's dynamic fidelity. The model not only captures the peak stress evolution and unloading slopes across cycles, but also preserves the internal dissipation pattern evident in the shaded areas between loading and unloading paths. The portion of the first ($G_2$) and second ($G_3$) deviatoric branch contributions highlights the viscoelastic structure of the model. The relatively small contribution of the first deviatoric branch (G2) compared to the second (G3) is attributed to the inherent material behavior, where long-term viscoelastic and damage-related responses dominate under cyclic loading.

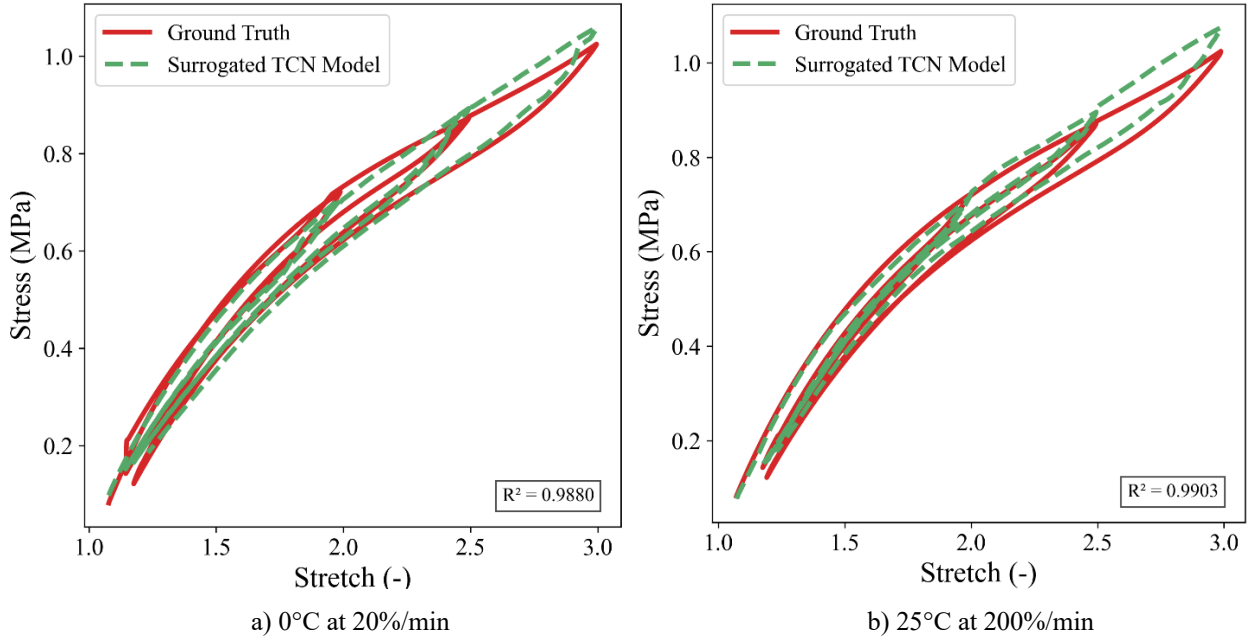

a) 0°C at 20%/min       b) 25°C at 200%/min

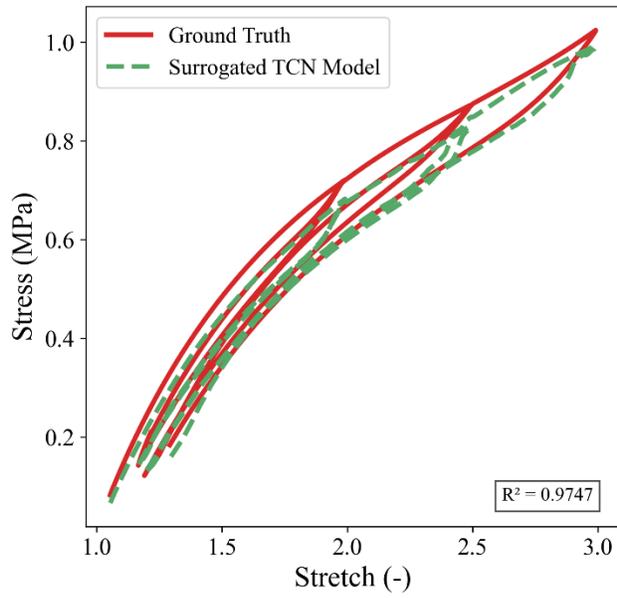
c) 55°C at 20%/min

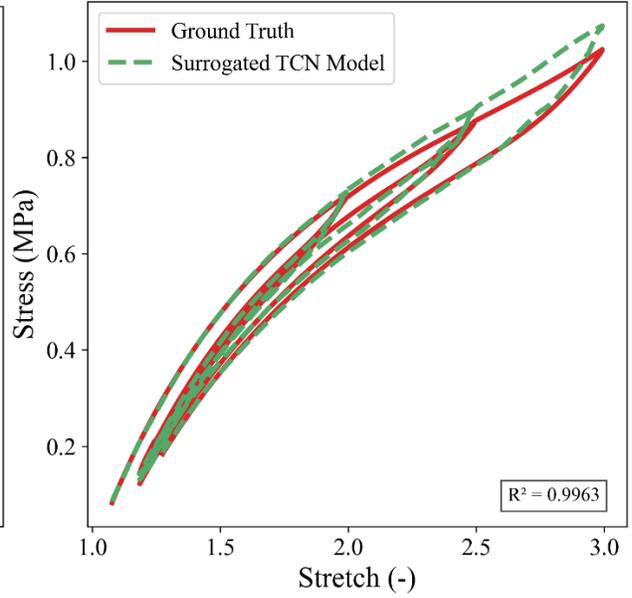
d) 55°C at 500%/min

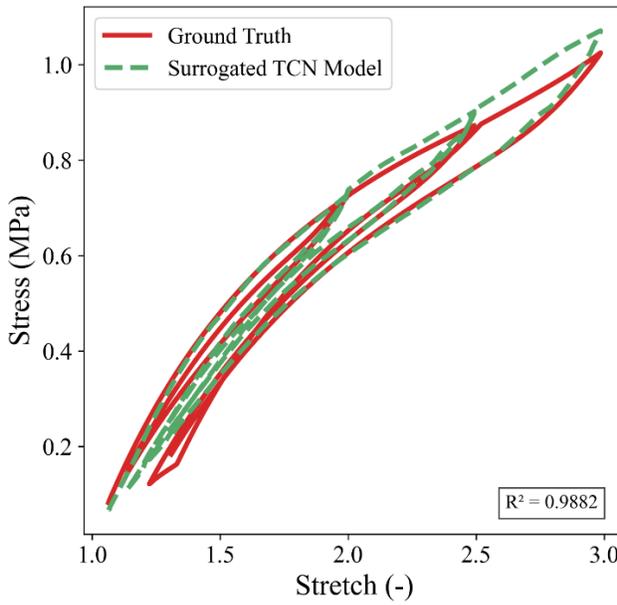
e) 25°C at 500%/min

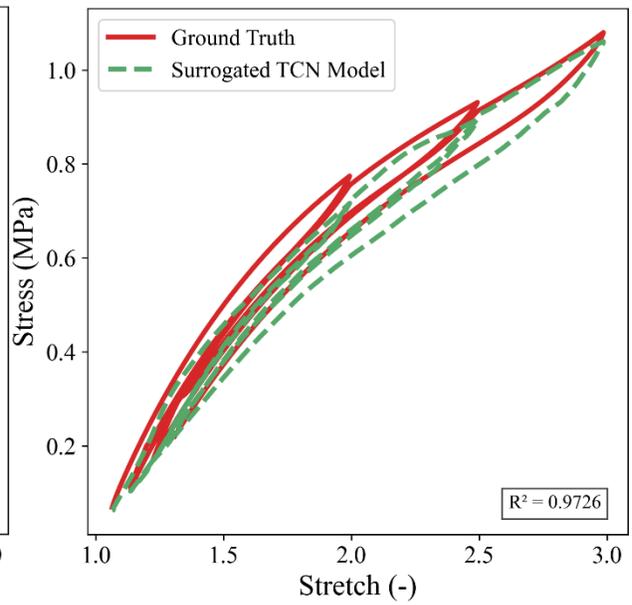
f) 10°C at 500%/min

Figure 7: Predicted vs. ground truth stress under varying temperatures and strain rates

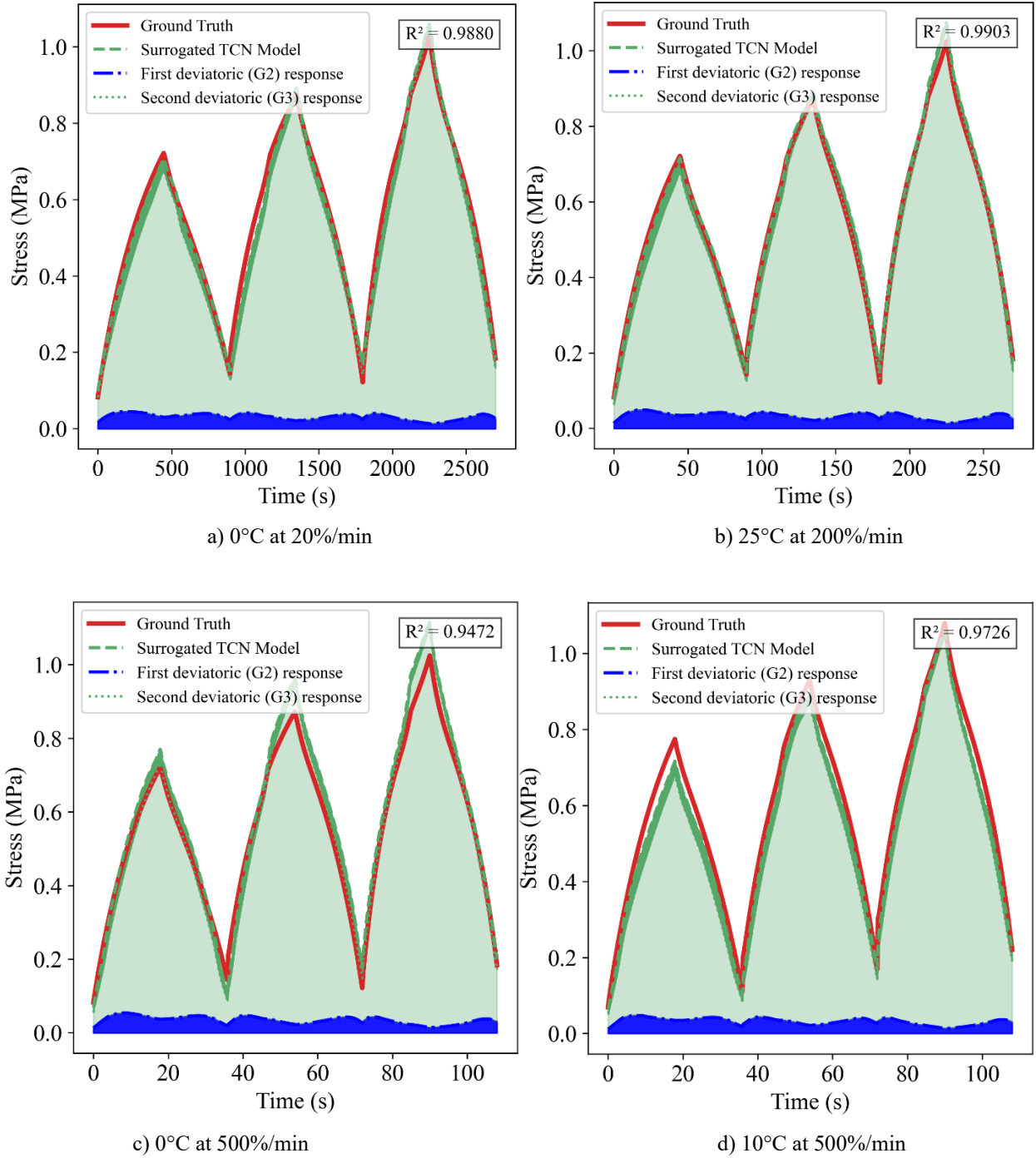

Figure 8: Surrogate model predictions and stress decomposition under multiple thermomechanical conditions

### 4.3.3 Case 3: Thermodynamic Consistency and Noise Robustness

To assess the thermodynamic consistency of the surrogate model, we evaluated the dissipation energy density over time under nine distinct thermo-mechanical loading conditions, as shown in **Figure 9 (a)–(c)**. The dissipation increases monotonically with both cycle count and stretch amplitude, which is consistent with physical expectations for viscoelastic soft materials under large deformation. Notably, the peak dissipation values scale with both strain rate and temperature. At higher strain rates, sharper and larger spikes in dissipation are observed at peak deformation, reflecting greater energy loss due to faster loading. In all cases, the model ensures non-negative dissipation, a key requirement for thermodynamic admissibility. This confirms that the surrogate model adheres to the second law of thermodynamics, reliably capturing internal energy loss mechanisms under diverse loading paths.

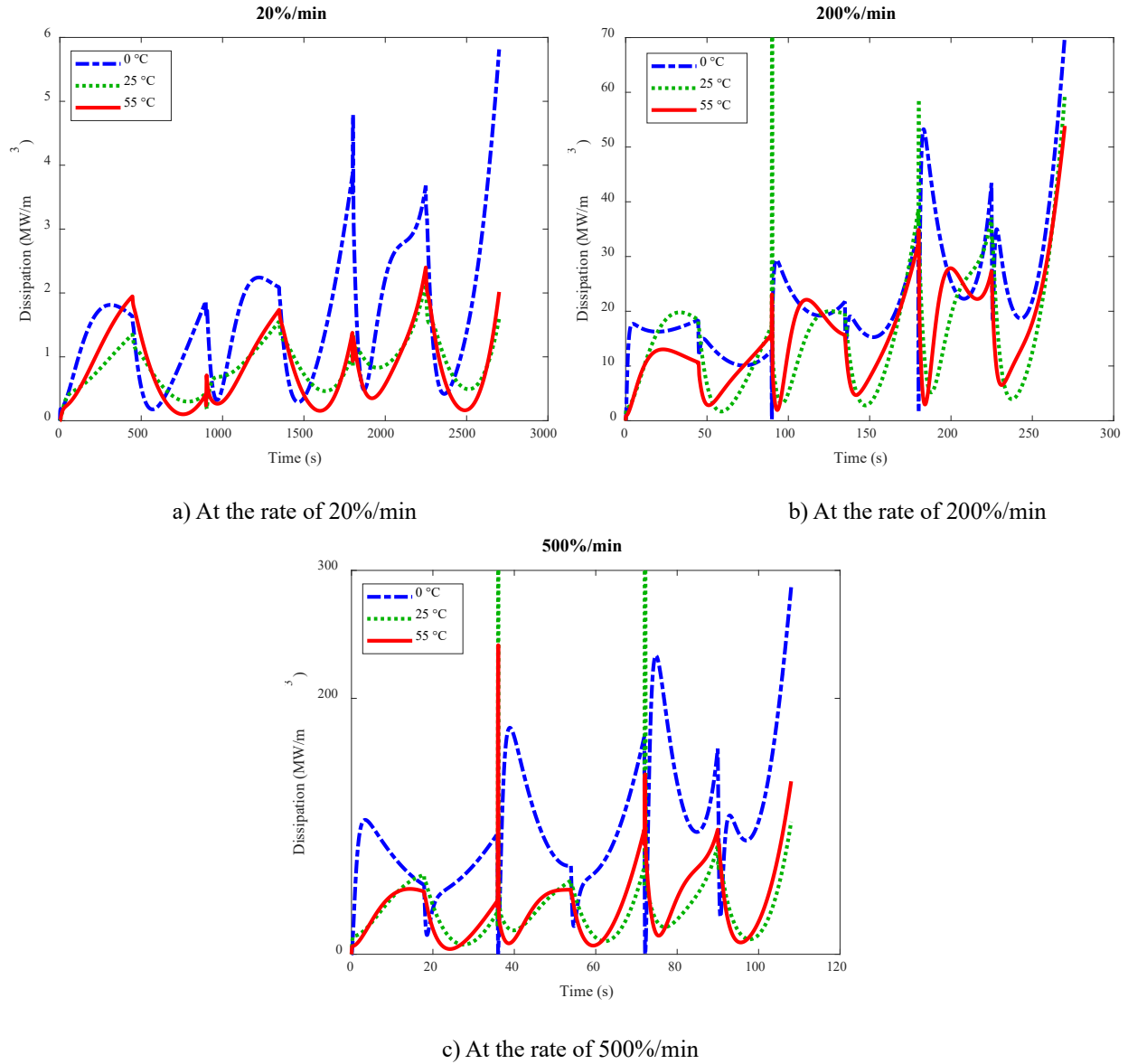

a) At the rate of 20%/min

b) At the rate of 200%/min

c) At the rate of 500%/min

Figure 9: Predicted dissipation energy evolution under nine thermo-mechanical conditions

To further evaluate the uncertainty-handling capability of the proposed surrogate model, we introduced Gaussian noise at a 20% level into the input stress data across multiple thermo-mechanical test conditions. This noise emulates experimental inconsistencies such as sensor fluctuations or environmental disturbances. The model was trained and evaluated on both clean and noisy data to quantify its performance. **Figure 10 (a)–(h)** presents the stress–stretch responses under varying combinations of temperature and strain rates. Despite the relatively high noise level, the TCN model trained on noisy data maintained excellent agreement with the corrupted ground truth trends. It successfully captured both the nonlinear stress–stretch responses, damage evolution, and loading–unloading hysteresis loops, even under challenging high-rate or low-temperature conditions. The predictions exhibited only minor local deviations, especially in the transition zones near loading reversals. This suggests that the model structure itself possesses strong inherent regularization, likely due to the embedded physics and history-dependent sequence encoding.

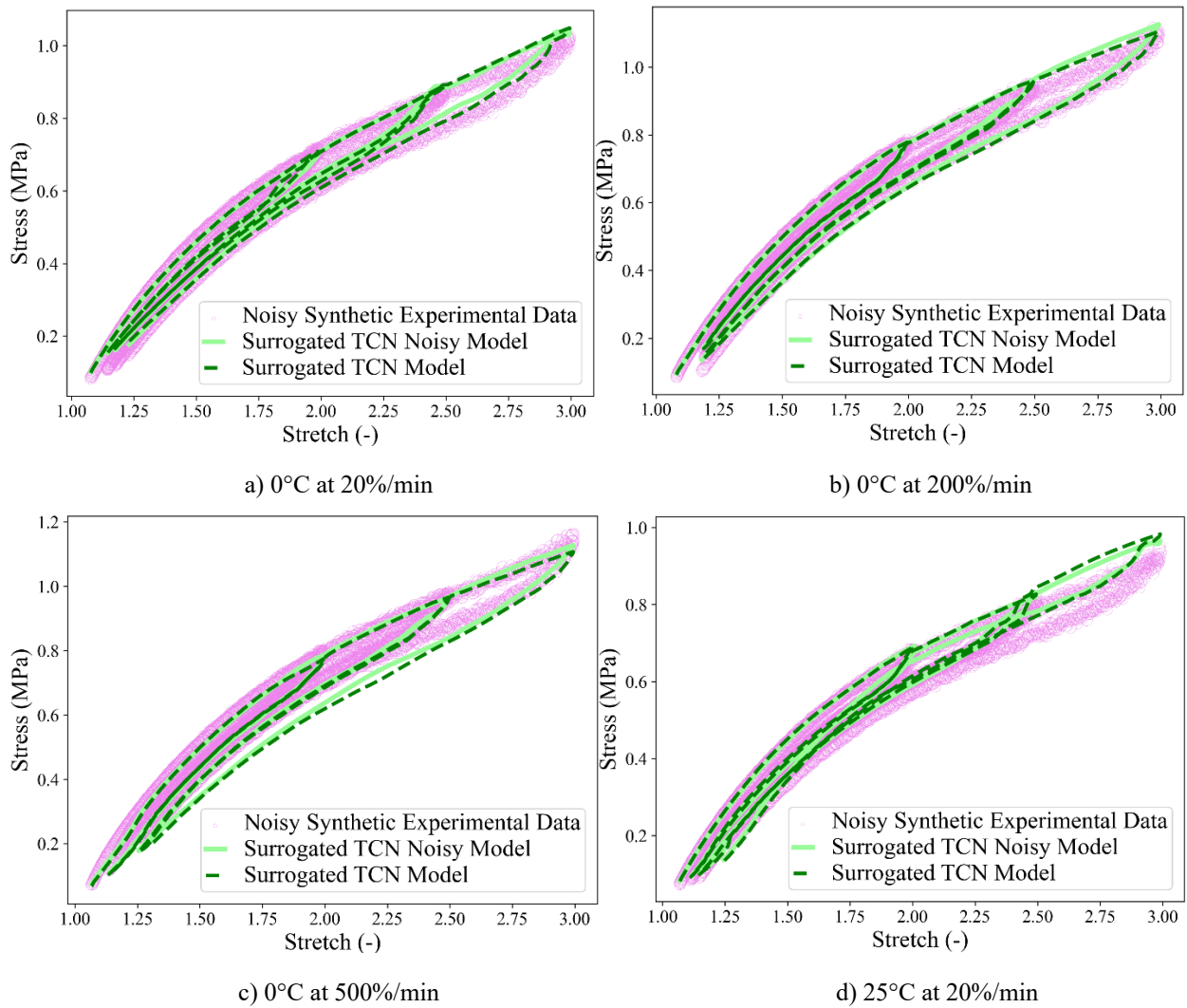

a) 0°C at 20%/min

b) 0°C at 200%/min

c) 0°C at 500%/min

d) 25°C at 20%/min

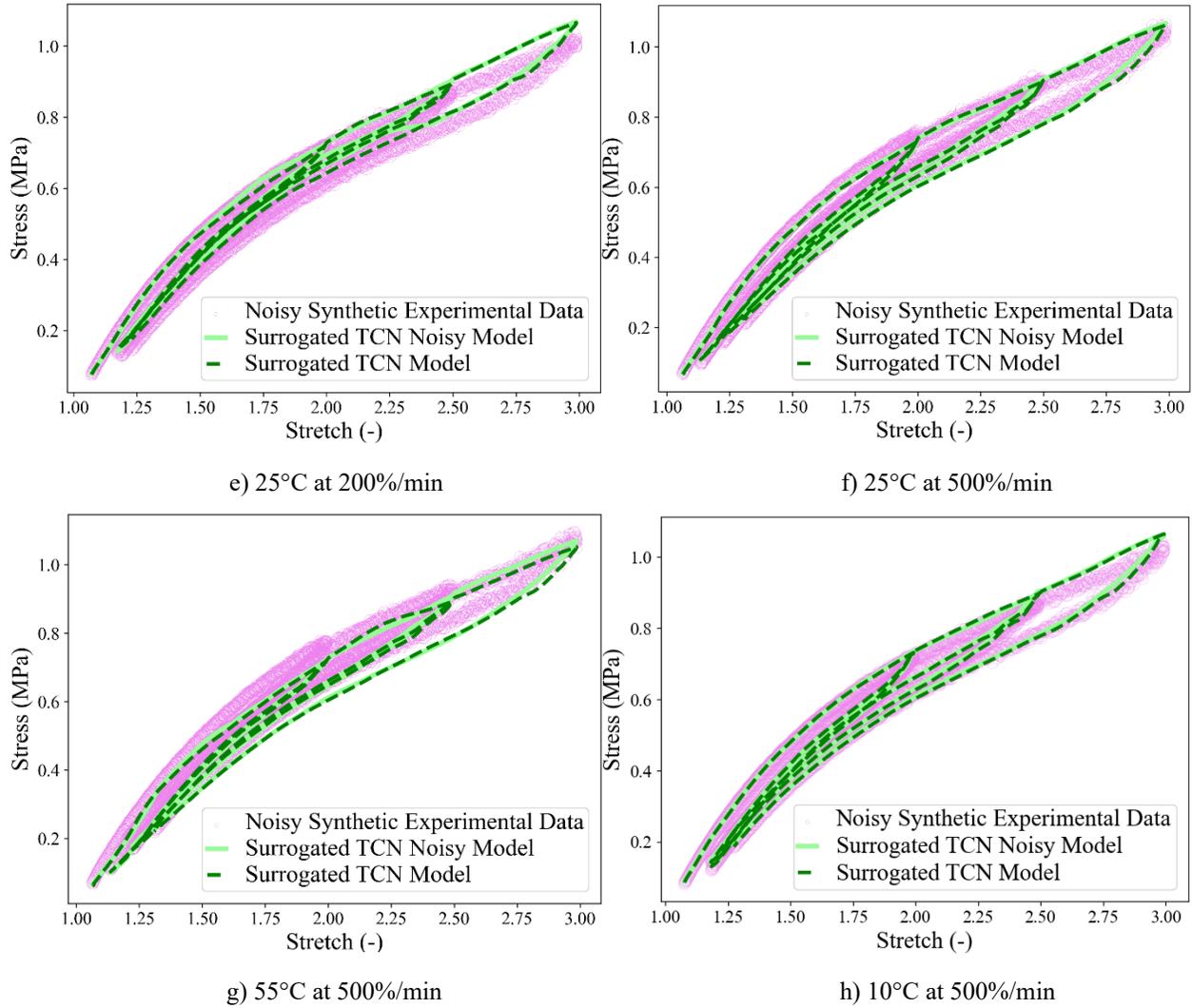

e) 25°C at 200%/min        f) 25°C at 500%/min

g) 55°C at 500%/min        h) 10°C at 500%/min

Figure 10: Surrogate model performance under 20% Gaussian noise

*4.3.4    Case 4: Damage Evolution in an Open-Hole Sample under High-Temperature Cyclic Uniaxial Loading*

We consider a plane-strain, open-hole square specimen subjected to uniaxial cyclic loading to investigate Mullins damage evolution. The geometry consists of a 2 × 2 mm square plate with a central circular hole of radius 0.5 mm. Due to symmetry, only the upper-right quarter of the domain is modeled (**Figure 11**). A series of three cyclic loading simulations is performed at progressively increasing stretch levels. All simulations are conducted at a fixed temperature of 55°C and a constant strain rate of 40%/min.

To simulate the material behavior, we use our group VUMAT subroutine in Abaqus/Explicit that captures finite strain viscoelasticity and Mullins damage evolution. The surrogate model takes stretch-based features derived from the deformation gradient as input and predicts the corresponding stress. These predictions are evaluated separately from the Abaqus simulation, allowing for comparison against the Abaqus reference solution. As seen in **Figures 12a-d** and **12e-h**, both methods predict Mullins damage growth with increasing stretch. At lower stretches, damage is localized near the hole, whereas at higher levels (50% stretch), damage extends across the specimen. The final damage saturates at approximately 0.09, with the TCN model closely replicating the Abaqus reference solution.

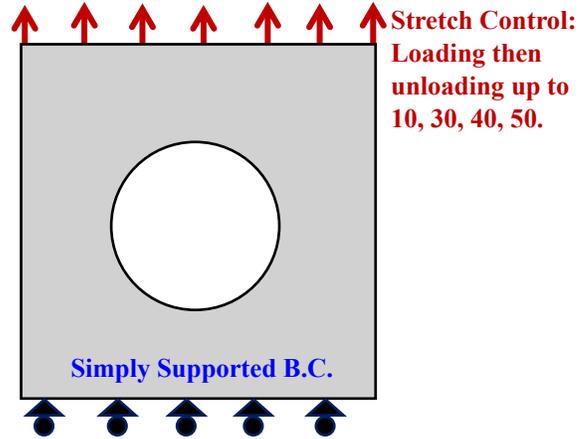

Figure 11: Schematic and boundary conditions of the open-hole sample under cyclic uniaxial test at the rubbery state 55°C

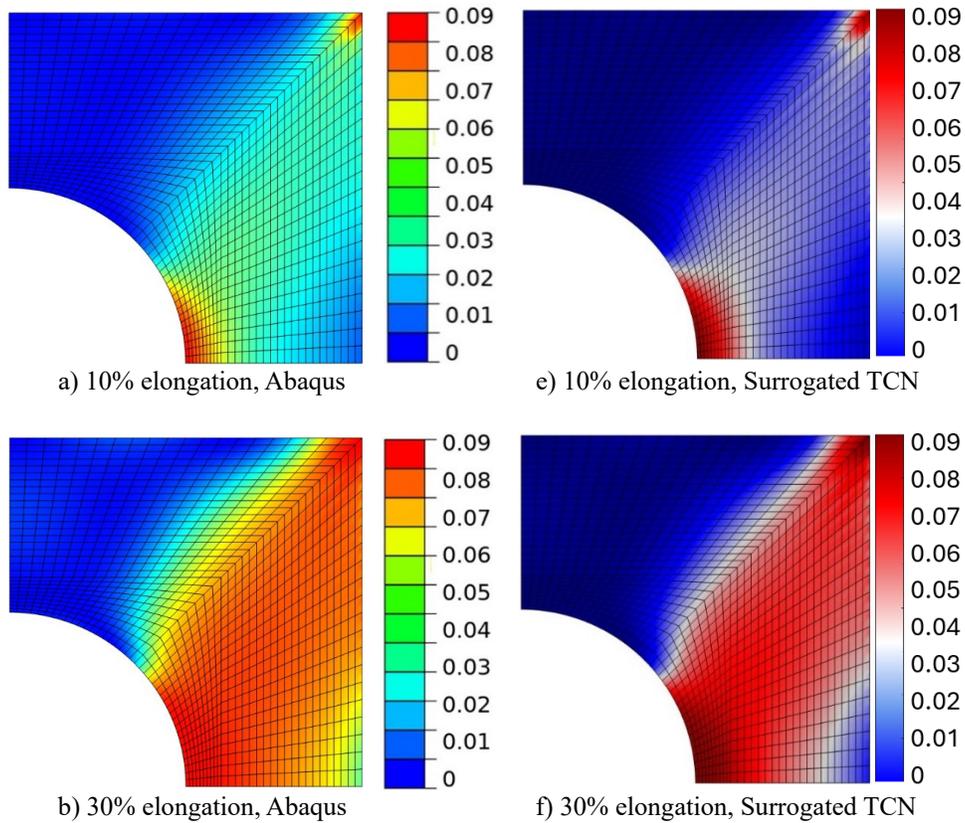

a) 10% elongation, Abaqus

e) 10% elongation, Surrogated TCN

b) 30% elongation, Abaqus

f) 30% elongation, Surrogated TCN

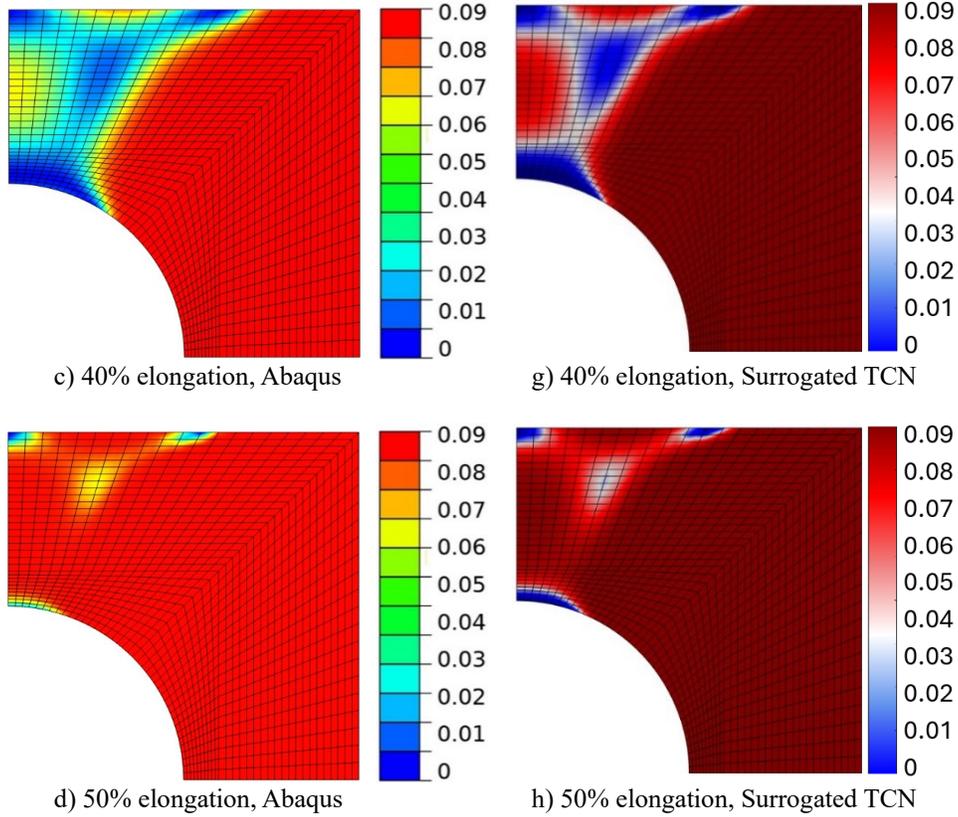

Figure 12: Damage contour plots in an open-hole specimen under increasing elongation levels

## 5. Conclusions

In this work, we developed a novel physics-informed, data-driven modeling framework that captures the complex interplay among time dependence, temperature effects, Mullins softening, and strongly nonlinear viscoelastic behavior in soft materials. The surrogate model, built on a TCN, is trained directly on ground truth data and rigorously evaluated across multiple axes of generalization, including entirely unseen stretch levels, untrained temperatures, and out-of-distribution strain rates.

The training dataset includes multi-cycle tests spanning three temperatures (0°C, 25°C, and 55°C), two strain rates (20 and 200%/min), and cyclic stretch levels reaching up to 150%, enabling the model to learn rich thermomechanical coupling and history-dependent stress responses. The model not only performs accurately within its training domain but also exhibits remarkable extrapolation capabilities, successfully predicting responses under unseen temperatures (e.g., 10°C), strain rates 2.5× higher than training (from 200%/min to 500%/min), and 50% increase in elongation (from 150% to 200%). These tests confirm the model's robustness in out-of-distribution regimes while maintaining physical realism, preserving dissipation trends, and capturing the correct viscoelastic structure.

Thermodynamic admissibility is enforced through a custom-designed loss function that incorporates the Clausius–Duhem inequality and imposes a bounded constraint on the damage parameter, ensuring physically consistent softening and dissipation. Crucially, the model incorporates the maximum isochoric invariant to track strain energy history, allowing it to accurately capture Mullins-type damage accumulation under repeated loading. The surrogate demonstrates robustness to input noise, accurately predicting loading–unloading hysteresis, internal dissipation, and viscoelastic memory effects across multiple cycles. To validate its generalization capabilities, the model was tested on several challenging conditions excluded from training. In all cases, it achieved high accuracy in reproducing nonlinear stress–stretch behavior, temperature-rate coupling, and energy dissipation. Stress–time plots reveal that the surrogate retains detailed temporal structure, correctly resolving viscoelastic branch contributions, with dominant response captured by the long-term deviatoric branch. Finally, the trained model was deployed in a finite element setting and tested on a plane-strain open-hole square specimen undergoing large cyclic deformation at 55°C. Using VUMAT-generated FEM results as reference, the TCN surrogate closely matched the stress response and successfully reproduced the spatial and temporal evolution of Mullins damage, including localized softening and saturation at 50% stretch.

**CRediT authorship contribution statement**

**Alireza Ostadrahimi:** Conceptualization, Investigation, Methodology, Formal analysis, Data curation, Software, Visualization, Writing – original draft. **Amir Teimouri:** Methodology Data, curation. **Kshitiz Upadhyay:** Methodology, review & editing. **Guoqiang Li:** Methodology, Supervision, Funding acquisition, Writing – review & editing.

**Declaration of competing interest**

The authors declare that they have no known competing financial interests or personal relationships that could have appeared to influence the work reported in this paper.

**Data availability**

Data will be made available on request.

**Acknowledgment**

The authors acknowledge the support of the US National Science Foundation under Grant Numbers OIA-1946231 and 2331294 as well as the Louisiana Board of Regents for the Louisiana Materials Design Alliance (LAMDA).